\documentclass{iopart}
\usepackage{graphicx}
\usepackage[hypertex]{hyperref}
\usepackage{iopams}
\renewcommand{\Im}{\mathrm{Im}}
\renewcommand{\vec}[1]{\mathbf{#1}}
\newcommand{\vect}[1]{\mathbf{#1}}
\newcommand{\ex}{\mathrm{ex}}

\begin{document}

\topical{Collective coherence in planar semiconductor microcavities}
\author{
J.~Keeling$^1$,
F.~M.~Marchetti$^2$
M.~H.~Szyma\'{n}ska$^3$, 
P.~B.~Littlewood$^1$}
\address{$^1$ Cavendish Laboratory, University of Cambridge, Madingley
Road, Cambridge CB3 0HE, UK} 
\address{$^2$ Rudolf Peierls Centre for Theoretical Physics,
  University of Oxford, 1 Keble Road, Oxford OX1 3NP, UK}
\address{$^3$ Clarendon Laboratory, Department of Physics, University
of Oxford, Parks Road, Oxford, OX1 3PU, UK} 

\date{\today}
\begin{abstract}
  Semiconductor microcavities, in which strong coupling of excitons to
  confined photon modes leads to the formation of
  exciton-polariton modes, have increasingly become a focus for the
  study of spontaneous coherence, lasing, and condensation in solid
  state systems.  
  This review discusses the significant experimental progress to date,
  the phenomena associated with coherence which have been observed,
  and also discusses in some detail the different theoretical models
  that have been used to study such systems.
  We consider both the case of non-resonant pumping, in which
  coherence may spontaneously arise, and the related topics of
  resonant pumping, and the optical parametric oscillator.
\end{abstract}


\section{Introduction}
\label{sec:introduction}

Semiconductor microcavities have been designed to greatly enhance the
matter-light interaction strength by confining light.
The confined light couples to excitonic resonances in the medium
inside the microcavity; when this exciton-photon coupling exceeds the
exciton and photon damping rates one finds spectrally separated normal
modes, microcavity exciton-polaritons~\cite{bjork91,weisbuch92}.
Unlike many other examples of strong coupling, studied in quantum
optics, this review will focus on planar semiconductor microcavities that
confine photons in only one direction, thus leading to a continuum of
strongly coupled modes.
Due to their dual matter-light nature, exciton polaritons can be
manipulated and studied through their light component, and have an
effective interaction through their matter component and the
nonlinearity of light-matter coupling.
Due to the continuum of modes, the behaviour of exciton polaritons may
be related to the statistical mechanics of interacting bosons.
Thus, semiconductor microcavities provide an ideal system in which to
study the interface between quantum optics, strong coupling,
spontaneous coherence and quantum condensation.

A closely related area of research, although one we will not address
in this review, is strong coupling to single excitonic resonances i.e.\
excitons in quantum dots in semiconductor microcavities.
As well as experiments on quantum dots in planar
microcavities~\cite{ramon06}, experiments with confinement of photons
in either three or two spatial directions are also
performed.
In the first case, 0D microcavities have been constructed from
photonic crystals~\cite{yoshi04,antonio05} (in which in-plane
confinement results from localisation on a defect in the photonic
crystal, and vertical confinement from total internal reflection), as
well as micropillars~\cite{reithmaler04} (where Bragg mirrors provide
vertical confinement, and total internal reflection provides in-plane
confinement).
In the second case, wire structures were fabricated by chemical
etching~\cite{dasbach2005}.
Lying between confinement in one and three spatial directions are
experiments in patterned microcavities, where schematically a
variation in the width of the microcavity provides a shallow in-plane
trap for photon modes~\cite{kaitouni06,daif06,daif06:PSS}.
This results in coexistence of 0D and 2D polariton states, separated
in energies: A clear polariton spectrum has been seen along with the
quantisation induced by a box-like confinement.
Polariton localisation due to the intrinsic photonic disorder has also
been observed~\cite{langbein02}.

Our review will concentrate on macroscopic collective phenomena
arising from the interaction between these special bosonic particles.
While some of these issues have been addressed in other contexts ---
such as quantum condensation in dilute atomic
gases~\cite{leggett01,pitaevskii03}, and coherent quantum optics of
lasers~\cite{haken70} --- the combination of effects seen in
microcavity polaritons calls for new approaches.
In part, the new theoretical challenges arise from description of
features of semiconductor microcavities, such as the disorder and
decoherence that arise in solids, the spin structure of polaritons,
and  the effect of pumping and decay.
In addition, polariton systems provide new opportunities for
experimental probes and observations that differ from those possible
in other systems.

The recent experimental and theoretical research in this field 
can be divided into two main directions:
Firstly, experiments which use resonant (coherent) pumping have been
motivated by the search for all-optical ultrafast switches and
amplifiers.
The other direction is that of non-resonantly pumped microcavities,
where experiments have pursued the search for Bose-Einstein
condensation (BEC), polariton lasing and macroscopic phase coherence
phenomena.

A number of reviews have been already written on the subject of
microcavity polaritons: In Refs.~\cite{skolnick98,savona99}, linear
properties of microcavities have been analysed in great
detail.
Refs.~\cite{khitrova99,ciuti03} review problems of
non-linear optics and the theoretical framework necessary to study
resonant pumping, parametric amplification and oscillation.
In addition, two books~\cite{yamamoto00,kavokin-book} and two special
issues~\cite{special1, special2} have also been published.
In this review, we will focus attention on those issues of modelling
microcavity polaritons that are of particular importance in
understanding spontaneous coherence and condensation in such systems.
We will address the relation between the various theoretical
approaches that have been used, and discuss the limits in which
they become equivalent.
We will also discuss in some detail the relation between coherence,
condensation, lasing and superfluidity in experimental systems which
are finite, two-dimensional, decaying, and interacting, and thus
differ from the Bose-Einstein condensation of ideal three-dimensional
bosons.
Finally, we will discuss the relation of these features both to the
resonantly pumped polariton system, and also to other experimental
systems in which similar issues of coherence and condensation in
complex systems are addressed.

The review is divided overall into
non-resonant pumping in Sec.~\ref{sec:non-resonant-pumping} 
and resonant pumping in Sec.~\ref{sec:resonant-pumping}.
Within Sec.~\ref{sec:non-resonant-pumping}, we first present the
experimental development of the subject in
Sec.~\ref{sec:summary-experiments-nonres}, and then discuss the
theoretical approach, dividing our discussion into the question of
choice of model in Sec.~\ref{sec:theoretical-models}, choice of
treatment (i.e.\ thermal equilibrium, rate equations, etc.)  in
Sec.~\ref{sec:theor-treatm-effects}, and the phenomena predicted (i.e.\
experimental signatures, conditions for condensation) in
Sec.~\ref{sec:phenomena}.
Within Sec.~\ref{sec:resonant-pumping} we again divide into a summary
of experimental progress in Sec.~\ref{sec:summary-experiments-res},
and a discussion of the additional theoretical issues relevant
only to the resonantly pumped case in Sec.~\ref{sec:theories}.
Section~\ref{sec:conn-other-syst} finally draws comparisons to
phenomena seen in other experimental systems, and briefly summarises
our discussion.

\subsection{Introduction to microcavity polaritons}
\label{sec:intr-micr-polar}

\begin{figure}
  \centering
  \includegraphics[width=0.7\linewidth]{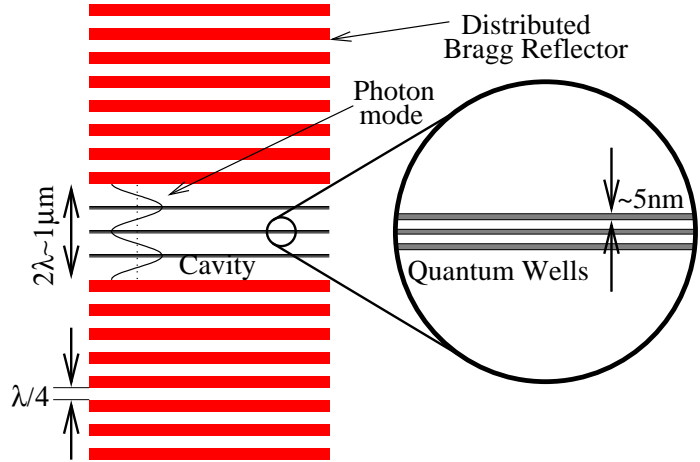}
  \caption{Schematic diagram of a microcavity, formed by a pair of
  distributed Bragg reflector stacks, with quantum wells at the
  antinodes of the cavity photon mode.}
  \label{fig:microcavity}
\end{figure}

Before discussing experiments and theories of coherence in microcavity
polaritons, we provide here a brief introduction to the systems
considered, and to microcavity polariton modes.
Fuller introductions can be found
elsewhere~\cite{yamamoto,yamamoto00,kavokin-book}.
The semiconductor microcavities we discuss are constructed from
distributed Bragg reflectors, containing alternating quarter
wavelength thick layers of dielectrics with differing refractive
indices.
Due to these Bragg reflectors, the cavity contains a standing wave
pattern of confined radiation.
As illustrated in Fig.~\ref{fig:microcavity}, quantum wells (QWs) are
placed at the antinodes of this standing wave, thus maximising the
coupling between photons and excitons confined to the quantum wells.

\begin{figure}
  \begin{center}
    \includegraphics[width=0.7\linewidth]{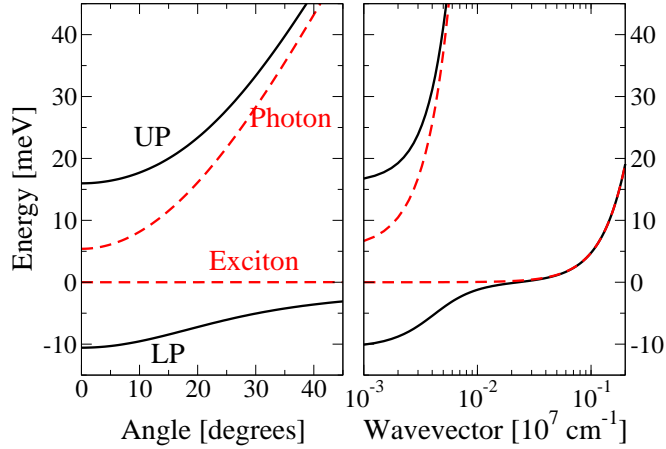}
  \end{center} 
  \caption{Schematic polariton spectrum. Left as a function of
    emission angle, $\theta=\sin^{-1}(ck/\omega_0)$; Right, as a
    function of momentum on a logarithmic scale, showing
    the full exciton dispersion. (Plotted for $M= 0.08 m_e$,
    $m=3\times10^{-5} m_e$, $\Omega_R=26$meV, $\delta=5.4$meV and
    $\omega_0=1.7$eV.)  }
  \label{fig:polariton}
\end{figure}

Because the photon modes are confined to the cavity, the volume
associated with the radiation mode is small, and so the exciton-photon
coupling is strong.
This strong coupling means that rather than considering the
exciton-photon coupling as leading to radiative decay of the excitons,
the exciton and photon modes are instead mixed, to form new normal
modes: lower and upper polaritons.
At the simplest level, one can write the exciton-photon Hamiltonian in
terms of operators $\psi_{\vec{k}}^{\dagger}$ creating photons and
$D_{\vec{k}}^{\dagger}$ creating excitons, with $\vec{k}$ labelling
the 2D in-plane momentum.
Thus:
\begin{equation}
  \label{eq:exciton-photon-quadratic}
  H = 
  \left( \psi_\vec{k}^{\dagger},  D^{\dagger}_{\vec{k}}\right)
  \left(
    \begin{array}{cc}
      \omega_{\vec{k}} & \Omega_R/2\\
      \Omega_R/2 & \varepsilon_{\vec{k}}
    \end{array}
  \right)
  \left( 
    \begin{array}{c}
      \psi_\vec{k}^{}\\
      D_{\vec{k}^{}}
    \end{array}
  \right).
\end{equation}
[We have set $\hbar=1$ here and throughout.]
Here, $\omega_{\vec{k}}$ is the energy of the photon mode confined in
the cavity of width $w$, giving: $\omega_{\vec{k}} = (c/n) \sqrt{k^2 +
  (2\pi N/w)^2}$,
with $n$ is the refractive index, and $N$ the index of the transverse
mode in the cavity.
For the situation in Fig.~\ref{fig:microcavity}, $N=2$.
For small $k$, the energy can be written as $\omega_{\vec{k}} = \omega_0 +
k^2/{2m}$, where $m$ is an effective photon mass $m = (n/c) (2\pi/w)$.
In the absence of disorder, the exciton energy in the QW is
$\varepsilon_{\vec{k}} = \varepsilon_0 + k^2/2M$, where $M$ is the
total exciton mass, and $\varepsilon_0 = E_{cv} - \mathcal{R}_{ex}$
comes from the conduction-valence band gap $E_{cv}$ including QW
confinement and the exciton binding energy (Rydberg)
$\mathcal{R}_{ex}$.
For convenience, we define the bottom of the exciton band,
$\varepsilon_0$ as the zero of energies; and denote the detuning between
exciton and photon bands as $\delta = \omega_0 - \varepsilon_0$.
Finally, the off diagonal term $\Omega_R/2$ describes the
exciton-photon coupling,  where $\Omega_R$ is the Rabi frequency.
Then, diagonalising the quadratic form in
Eq.~(\ref{eq:exciton-photon-quadratic}) gives the polariton spectrum:
\begin{equation}
  \label{eq:pol-spectrum}
  E_{\vec{k}}^{\mathrm{LP},\mathrm{UP}}
  =
  \frac{1}{2} \left[
    \left(\delta + \frac{k^2}{2M} + \frac{k^2}{2m} \right)
    \mp
    \sqrt{%
      \left(\delta + \frac{k^2}{2M} - \frac{k^2}{2m} \right)^2
      +
      \Omega_R^2
    }
  \right].
\end{equation}
This spectrum is illustrated in Fig.~\ref{fig:polariton}. 
It is shown there both as a function of momentum $k$, and also as a
function of angle.
The angle corresponds to the angle of emission of a photon out of the
cavity; since the in-plane momentum and photon frequency are both
conserved as photons escape through the Bragg mirrors, one may write
$(\varepsilon_0 +E^{LP}_{\vec{k}})\sin(\theta) = c k$, which (since
typical values of $k$ satisfy $\omega_0, \varepsilon_0 \gg c k$) can be
approximated as $\omega_0 \sin(\theta) = k^2/2m$.
%

\section{Non-resonant pumping}
\label{sec:non-resonant-pumping}

\subsection{Summary of experiments}
\label{sec:summary-experiments-nonres}

Optical properties of semiconductor microcavities have been the subject of
extensive experimental investigations since the first observation of
the strong coupling regime by Weisbuch {\it et al.}~\cite{weisbuch92}.
Much of the experimental research has concentrated on III-V materials,
mainly GaAs/AlGaAs structures, or on II-VI materials, such as
CdTe/CdMnTe/CdMgTe structures.
The main aim of the experiments described here as non-resonant pumping
has been to start with incoherently injected polaritons, 
and observe {\em spontaneous} coherent processes emerging
from incoherent injection of polaritons:
polariton degeneracy, final state stimulation, and ultimately
polariton BEC.

\begin{figure}[htbp]
  \centering
  \includegraphics[width=0.55\linewidth]{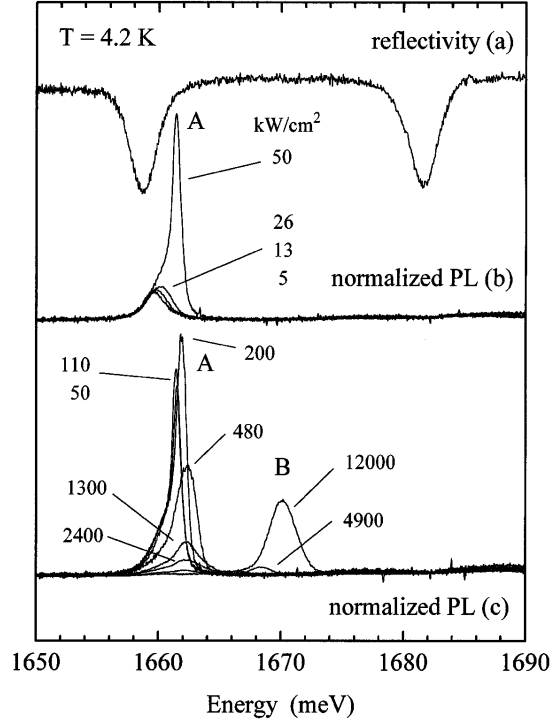}
  \caption{ Panel (a): Reflectivity spectrum, showing location of
    lower and upper polariton modes in the absence of pumping.  Panel
    (b): Photoluminescence as pumping power is increased, a threshold
    for nonlinear emission is seen at $40$kW/cm$^2$, while the system
    clearly remains in the strong coupling regime.  In panel (c), at
    much higher pumping powers ($10,000$kW/cm$^2$), a second
    threshold, to electron-hole lasing is seen.
    [From~\cite{dang98:prl} Copyright (1998) by the American Physical
    Society]}
  \label{fig:dang98emission}
\end{figure}

The authors of the earliest report~\cite{pau96} of non-linear emission
in the presence of strong coupling in GaAs microcavities, suggesting
final state stimulation characteristic for bosonic particles, later
withdrew those conclusions~\cite{cao97} as further experiments showed
that the threshold for non-linear emission occurred in that case after
the crossover to the weak-coupling regime; thus the non-linear emission
should have been attributed to photon lasing.
The first unambiguous observation of polariton bosonic stimulation
was in CdTe microcavities~\cite{dang98:prl} consisting of 16
quantum wells with a Rabi splitting of around $23$meV.
Two distinct stimulation thresholds were observed with increasing
intensity of continuous wave pumping, as shown in
Fig.~\ref{fig:dang98emission}.
As the pumping intensity was increased above the first threshold,
non-linear emission at energies close to the bottom of the lower
polariton branch was clearly seen.
The second threshold, reported for much higher intensities, was
connected with a weak coupling electron-hole lasing mechanism.
Further investigation~\cite{boeuf00,alex01} showed that the first
nonlinear threshold --- in the strong coupling regime --- was
due to stimulated scattering to the ground state.
Very shortly after publication of Ref.~\cite{dang98:prl}, non-linear
emission was seen in a single QW GaAs microcavity~\cite{sen99},
characterised by 3.5meV Rabi splitting.
However further investigation~\cite{sen00} showed that this
nonlinearity had emission varying as the square of pumping intensity,
and a threshold that occurred for occupation factors much less than
one, and so this nonlinearity was associated with increase of
exciton-exciton scattering, rather than final state stimulation.
Final state stimulation in III-V materials was demonstrated by
pump-probe experiments in Refs.~\cite{sen00, huang00}.

\begin{figure}[htbp]
  \centering
  \includegraphics[width=0.6\linewidth]{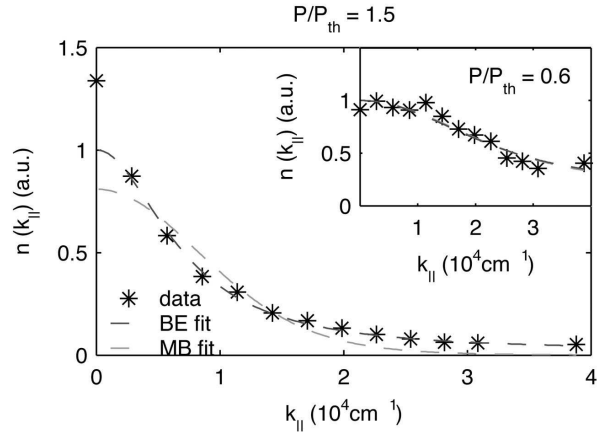}
  \caption{Momentum space distribution of lower polaritons above 
    (main figure) and below (inset) the nonlinear threshold. [From
    Ref.~\cite{deng03}, Copyright 2003 National Academy of Sciences]}
  \label{fig:deng03nk}
\end{figure}

The stimulated scattering to the ground state, and non-linear build-up
of lower polariton population was the first step towards demonstration
of spontaneous coherence and thermalisation --- characteristic of
quantum condensation.
However, the big challenge to realising a condensed polariton phase
was the finite (though very large) quality of the cavity mirrors, and
the resultant short polariton lifetime, of the order of picoseconds.
In addition, due to the `bottleneck effect',~\cite{tassone97:prb} the
relaxation of polaritons to the zero momentum state was delayed,
hindering the creation of a thermal population in the lowest energy
states.
The first investigation of the coherence properties of emitted light
above the threshold for non-linear emission in strong coupling was
based on the measurement~\cite{deng02,weihs03} of the second order
coherence function, $g_2(t=0)$, which would take a value of
$g_2(0)=2$ for a thermal state, and $g_2(0)=1$ for a coherent
state~\cite{gardiner}.
A decrease of $g_2(0)$ from $1.8$ to $1.4$ as pumping power was increased
from threshold to $20$ times threshold power was seen in a system of
$12$ GaAs quantum wells placed at the antinodes of light in a GaAs/AlGaAs
microcavity, giving $14.9$meV Rabi splitting.
This was followed by a report of a characteristic change in the
momentum space distribution above threshold~\cite{deng03}, as shown in
Fig.~\ref{fig:deng03nk}, and a blueshift 
of the polariton dispersion~\cite{weihs04:_polar}.
Time-resolved photoluminescence measurements were also performed for a
CdTe microcavity with $10.5$meV Rabi splitting~\cite{bloch05} under
non-resonant pulsed excitation, which were able to monitor the
buildup of a large polariton population in the $\vec{k}=0$ state.
Analysis of the time dependence showed that, below threshold, the
dynamics of the $\vec{k}=0$ polaritons follows closely the population
of cold reservoir excitons; the relaxation from high energy
exciton states resonant with the pump to these reservoir exciton
states had a characteristic relaxation time of $30$ps.
[Note that this time is significantly shorter than the $150$ps
observed in similar experiments~\cite{bloch02} with GaAs based
microcavities.]
Above the non-linear threshold, the population of the reservoir
excitons was found to be clamped, and the polariton relaxation
dynamics became faster, with the maximum of polariton emission at
$70$ps delay after the initial pulse.
This delay was further decreased for higher excitations powers.
Together these provide evidence of stimulated exciton-exciton
scattering to the lower polariton states.

The first evidence of spontaneous first-order coherence in an
incoherently pumped microcavity was seen in a 16 QW CdTe microcavity
with 26meV Rabi splitting~\cite{richard05} under non-resonant pulsed
pumping.
An interesting feature of this particular experiment was that the
non-linear emission was at $\vec{k} \neq 0$ and so resulted in an
emission ring at an angle of around $17 ^{\circ}$; this was associated
with the small size of the excitation spot (3 $\mu$m).
(Note that in later experiments with larger excitation spots on the
same sample condensation was at $\vec{k}=0$~\cite{kasprzak06:nature}.)
The first-order coherence was investigated by spectroscopic imaging
of the far-field emission.
Two momentum space images were superimposed giving fringes (as a
function of momentum $\vec{k}$) with over 75$\%$ contrast above threshold
and up to 35$\%$ below threshold.
In a later publication of the same group~\cite{richard05:prb},
experiments on a 4 QW CdTe microcavity characterised by $13.2$meV
Rabi splitting showed macroscopic occupation of the $\vec{k}=0$ state
characterised by narrowing above threshold of the polariton emission
line to a linewidth below that of the cavity photon mode.
Near-field images showed modulation of the polariton spatial
distribution, revealing the effect of photonic disorder.
The next challenge in the search for spontaneous condensation was to
see similar effects, but accompanied by a thermal distribution of
$\vec{k} \neq 0$ polaritons.

\begin{figure}[htbp]
  \centering
  \includegraphics[width=0.6\linewidth]{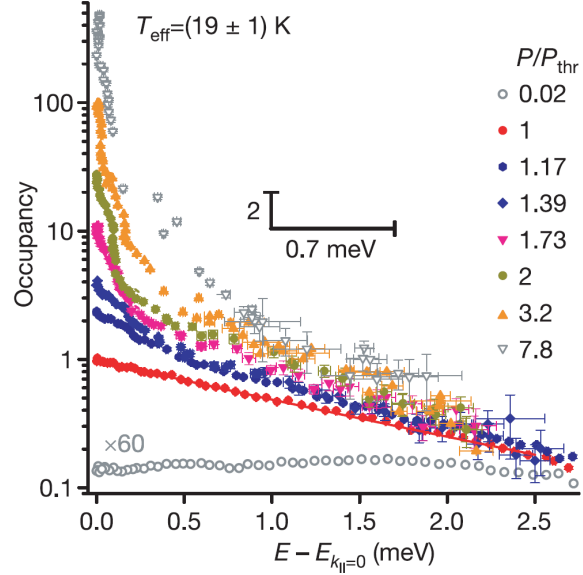}
  \caption{Polariton occupation vs energy, showing evolution from below
    to above threshold power, with the distribution remaining
    thermalised at a temperature of $19K$ [From
    Ref.~\cite{kasprzak06:nature}]}
  \label{fig:kasprzak06occ}
\end{figure}

Due to the short polariton lifetime and the `bottleneck
effect',~\cite{tassone97:prb} the realisation of equilibrium
population has proven to be challenging.
Progress came from observing~\cite{deng06:eqbm,kasprzak06:nature} that
thermalisation processes due to particle-particle scattering can be
dramatically increased both by increasing the value of the
(non-resonant) pump power, and also by positively detuning the cavity
energy above the excitonic energy.
Large positive detuning makes polaritons more excitonic and increases
their scattering rate.
Time- and angle-resolved spectroscopy on a sample consisting of 12 GaAs
QWs characterised by 14.4meV Rabi splitting~\cite{deng06:eqbm} showed
that for positively detuned cases, where the thermalisation time
increases while decay time decreases, the thermalisation time can
reach around one tenth of the polariton lifetime and that lower
polaritons remain in thermal equilibrium with the phonon bath for a
period of about 20ps.
Finally, a comprehensive set of experiments showing clear evidence
for condensation of cavity polaritons was performed in a
CdTe~\cite{kasprzak06:nature} structure consisting of 16 quantum
wells giving 26meV Rabi splitting.
Above the threshold pumping density they observed:
a massive occupation of the $\vec{k} \cong 0 $ mode developing from a
polariton gas in thermal equilibrium at 19K (shown in
Fig.~\ref{fig:kasprzak06occ}); an increase of temporal coherence from
$1.5$ps below threshold to up to $6$ps above threshold; the build-up
of long-range spatial coherence over the whole system size with
contrast of interference fringes from less than 5$\%$ below threshold
to 45$\%$ above threshold; linear polarisation of the emission.
Linear polarisation had been predicted to appear in the condensed
state~\cite{laussy06}, and its appearance gives evidence for the
single state nature of the condensate, though since the direction of this
polarisation was pinned to a crystallographic direction the
polarisation direction symmetry was not spontaneously broken.
Evidence that the polarisation of light is pinned to one of the
crystallographic axes, independently of the excitation polarisation,
was also independently observed in experiments on both
CdTe~\cite{klopotowski2006} and InGaAs microcavities~\cite{amo2005}.
These results were ascribed to birefringence in the mirrors and
cavity.
Shortly following the work in Ref.~\cite{kasprzak06:nature}, a
similar non-linear build up accompanied by a linear polarisation was
seen~\cite{snoke06:condensation} in GaAs structures in stress-induced
traps~\cite{Snoke06}.
Also the second order coherence function, $g_2(t=0)$, has been
measured \cite{kasprzak06:thesis} in the CdTe structures studied in
Ref.~\cite{kasprzak06:nature} where, in contrast to observations
reported in Ref.~\cite{deng02}, $g_2(0)$ was found to be around 1 at
threshold and then increased up to around 1.4 at powers 10 times
threshold.
This effect has been attributed to the phase diffusion due to
interactions.

Wide-band-gap semiconductor structures based on group-III nitrides,
such as GaN based cavities have recently attracted considerable
interest (see,
e.g.,~\cite{sellers2006_a,sellers2006_b,savvidis06,butte06,christmann06,tawara04,lidzey99,lidzey98}).
The main advantage of these structures over II-VI and other III-V
materials lies in the large exciton binding energy (around $26$meV for
bulk structures, and over $40$meV for narrow quantum wells) and the
large coupling to the photon field, which makes them ideal systems for
the realisation of functional devices operating at room temperature.
Although the study of cavity polaritons in group-III nitrides
microcavities is still in its infancy, strong exciton-photon coupling
in a bulk GaN cavity~\cite{sellers2006_a,butte06,sellers2006_b} and
in a quantum well cavity~\cite{christmann06} have been reported.
In both cases, the substantial inhomogeneous broadening of the
excitonic and photonic lines play a key role in establishing the
conditions for reaching strong coupling.
Very recently, a non-linear build-up of polariton emission accompanied
by an increase in the first-order temporal coherence, and a
spontaneously chosen linear polarisation (independent of the
apparatus and different between measurements) has been reported to
occur at room temperature in GaN bulk microcavities~\cite{baumberg06}.

\subsection{Theoretical models}
\label{sec:theoretical-models}

In this section, we will discuss the different models that have been
used to describe microcavity polaritons and study their condensation.
We wish to separate clearly two aspects of theoretical description of
polaritons; the first aspect is the choice of model, the subject of
this section, the second aspect is how that model is treated, which
will instead be covered in Sec.~\ref{sec:theor-treatm-effects}.
After having addressed these points, we then in
Sec.~\ref{sec:phenomena} discuss the various theoretical predictions
of conditions for condensation, and of possible signatures.
Readers who are not interested in the details of how the system is
theoretically modelled should jump to Sec.~\ref{sec:phenomena}.
A model starting from electrons and holes, taking into account their
Coulomb interaction to form bound excitons, their coupling to light,
and the effects of disorder would describe polariton systems exactly,
but is too complicated to allow any clear understanding of the
important features associated with condensation to be gained.
Therefore, it is appropriate to use simplified models,
that exaggerate some features of the real system, and neglect others.
In judging which model is appropriate to address a given problem, it
is important to understand how the model relates to the underlying
microscopic model of electrons and holes, and so we shall start by
discussing this microscopic model.

\subsubsection{Microscopic electron-hole Hamiltonian}
\label{sec:micr-electr-hole}

In this section, we discuss the underlying description of microcavity
polaritons formed from photons confined to a two-dimensional cavity,
interacting with electrons and holes in two-dimensional quantum
wells~\cite{haug,yamamoto}.
\begin{equation}
  \label{eq:eh-photon-hamiltonian}
  H = H_{\mathrm{eh}} + H_{\mathrm{coul}} + H_{\mathrm{disorder}} +
  H_{\mathrm{photon}} + H_{\mathrm{dipole}}
\end{equation}

Consider first the electrons and holes, we have
\begin{eqnarray}
  \label{eq:eh-hamiltonian}
  H_{\mathrm{eh}} &=& \sum_\vect{k} \varepsilon^c_\vect{k}
  c^{\dagger}_\vect{k} c^{}_\vect{k} + \varepsilon^v_\vect{k}
  v^{\dagger}_\vect{k} v^{}_\vect{k} \\
  \label{eq:coul-hamiltonian}
  H_{\mathrm{coul}} &=& \frac{1}{2 A} \sum_\vect{q} \frac{e^2}{2
  \epsilon_r \epsilon_0 |\vec{q}|} \left[ \rho^e_\vect{q}
  \rho^e_{-\vect{q}} + \rho^h_\vect{q} \rho^h_{-\vect{q}} -2
  \rho^e_\vect{q} \rho^h_{-\vect{q}} \right] \\
  \label{eq:disorder-hamiltonian}
  H_{\mathrm{disorder}} &=& \int d\vect{r} \left[W_e(\vect{r})
  c^{\dagger}(\vect{r}) c^{}(\vect{r}) - W_h(\vect{r}) v^{}(\vect{r})
  v^{\dagger}(\vect{r})\right] \; .
\end{eqnarray}
Here $c^{\dagger}_\vect{k}$ ($v^{\dagger}_\vect{k}$) create electrons
in the conduction (valence) bands, which have dispersions,
$\varepsilon^{c}_\vect{k}$ ($\varepsilon^{v}_\vect{k}$).
Since the ``empty'' state is a filled valence band, it is more
convenient to describe the valence band via the operator
$v_{\vect{k}}$ which creates a hole --- i.e.\ a missing electron.
The density of electrons (holes) is given by $\rho^e_\vect{q} =
\sum_\vect{k} c^{\dagger}_{\vect{k}+\vect{q}} c^{}_{\vect{k}}$
($\rho^h_\vect{q} = \sum_\vect{k} v^{}_{\vect{k}}
v^{\dagger}_{\vect{k}+\vect{q}} $). 
The factor $1/A$, where $A$ is the quantisation area of the cavity,
appears explicitly because the Hamiltonian has been written as a sum
over momentum labels; this factor plays no role in any final answer,
and is absorbed in the definition of $d\vect{k}$ if summation is
replaced by integration.
Note also that in general there should be a dependence on the electron
and hole spin degrees of freedom, that we neglect here.
The last term, Eq.~(\ref{eq:disorder-hamiltonian}), describes the
disorder potential acting on electrons and holes, e.g.\ due to
well-width fluctuations and alloy disorder.
In general, disorder can act differently on electrons and holes; in
practice for the materials used, the energy scale of disorder is less
than the binding energy, so disorder does not dissociate
excitons~\cite{zimmermann92}.
If the exciton binding energy is significantly larger than a
characteristic energy scale of disorder, then as described in
Ref.~\cite{zimmermann92} one can factorise the wavefunction into a
centre of mass wavefunction, and a wavefunction of relative
electron-hole separation.
Then, in the equation for the centre of mass wavefunction, one has an
effective disorder potential that is the result of convolving the
original disorder with the wavefunction for relative electron-hole
separation.
As a result of this convolution, the effective disorder potential as
seen by the exciton centre of mass wavefunction is smoothed over the
scale of the exciton Bohr radius~\cite{runge02}.

Turning now to the interaction with the photons,
\begin{eqnarray}
  \label{eq:photon-hamiltonian}
  H_{\mathrm{photon}} &=& \sum_{\vect{q}} \omega_\vect{q}
  \psi^{\dagger}_{\vect{q}} \psi^{}_{\vect{q}} + \int d\vect{r}
  W_{\mathrm{ph}}(\vect{r}) \psi^{\dagger}(\vect{r}) \psi(\vect{r}) \\
  \label{eq:dipole-hamiltonian}
  H_{\mathrm{dipole}} &=& \frac{1}{\sqrt{A}} \sum_{\vect{q},\vect{k}}
  e \mu_{cv} \sqrt{ \frac{\omega_\vect{q}}{2 \epsilon_r \epsilon_0
  L_w} } \left( \psi_{\vect{q}}^{\dagger}
  v^{\dagger}_{\vect{k}+\vect{q}} c^{}_{\vect{k}} + \mathrm{H.c.}
  \right)
\end{eqnarray}
In Eq.~(\ref{eq:dipole-hamiltonian}), the quantisation volume for the
electromagnetic field has been factored into $A L_w$, where $L_w$ is the
width of the cavity, and $A$ the quantisation area as discussed above.
The term $\mu_{cv}$ is the inter-band dipole matrix element, which can
be calculated given the Bloch wavefunctions of the two bands.
The term $W_{\mathrm{ph}}(\vect{r})$ in
Eq.~(\ref{eq:photon-hamiltonian}) describes photonic disorder, which
can arise due to roughness of the Bragg mirrors --- i.e.\ due to layer
width fluctuations (monolayer mismatch), or crystal
dislocations~\cite{langbein02,gurioli05}.
The effects of this photonic disorder, and of the exciton disorder
introduced above, can be quite different.
The photonic disorder is generally on large length scales (typically of
the order of a micrometre), comparable to the size of the excitation
spot, and so it is primarily associated with the spatial inhomogeneity
of polaritons seen in experiment~\cite{richard05:prb}.
In contrast, as discussed in Sec.~\ref{sec:boson-ferm-gener},
excitonic disorder is on much shorter length scales (typically of the
order of ten nanometres for CdTe), and thus does not affect the
spatial polariton density profile; however excitonic disorder does
have a significant impact on the distribution of excitonic oscillator
strengths.
Although the excitons are localised, in the absence of photonic
  disorder the polaritons formed consist of a superposition of many
  different localised excitons and extended photon states, and thus
  one may form delocalised polaritons from localised
  excitons~\cite{whittaker98}.
We will not explicitly discuss the effects of photonic disorder
further, however the discussion of condensation in a trap in
Sec.~\ref{sec:lasing-cond-superfl} can apply also to trapping in
disorder, as well as any deliberately engineered trapping.

The Hamiltonian in Eq.~(\ref{eq:eh-photon-hamiltonian}) already
contains a number of important approximations, which should be
discussed.
The interaction of photons with electrons and holes makes use of both
the dipole approximation, and the rotating wave
approximation~\cite[Chapter 10]{yamamoto}.
The interaction strength here is written in the dipole (length) gauge.
The choice between the dipole (length) gauge and the Coulomb
(velocity) gauge is not arbitrary, as the terms assigned as describing
free particles (without interaction with radiation) are different in
each gauge~\cite{starace71,kobe79:gauge,cohen-tannoudji89}.
This point is worth stressing, as the electromagnetic interaction
between excitons is split between the direct Coulomb term, and
a photon mediated term.
Thus the choice of gauge affects also the Coulomb interaction
[Eq.~(\ref{eq:coul-hamiltonian})], controlling which parts of it are
absorbed into the definition of exciton states, which parts are
associated with the ``photon'' operators --- in the Dipole gauge, the
fields $\psi^{\dagger}_\vect{q}$ are quantised modes of the electric
displacement --- and which should be written as some effective
exciton-exciton interaction~\cite{cohen-tannoudji89}.
The relation between Coulomb interaction and photon mediated
interaction is complicated here because the resonant photons are
confined by the DBR (distributed Bragg reflector) mirrors, while the
static Coulomb term is modified much less strongly by the mirrors.
When one comes to exciton states, it is therefore important to be
aware that the choice of gauge affects both the exciton-photon
coupling strength, and the form of the inter-exciton Coulomb
interaction, and that these two are not separate.

In the next two sections, we will discuss the main two classes of
effective Hamiltonians, derived from this full Hamiltonian, used
to study microcavity polaritons.
The differences between these effective Hamiltonians can be seen as
the result of regarding different terms as important; i.e.\ which
terms are treated exactly, and which perturbatively.
In both cases, the first step involves changing from electrons and
holes to bound excitons --- i.e.\ solving the wavefunction for the
relative coordinates.
The differences then arise from considering in one case next the
effect of disorder, giving localised states, and then approximating
the inter-exciton Coulomb term by exclusion --- this leads to the
boson-fermion model discussed in Sec.~\ref{sec:boson-ferm-gener} ---
or alternatively, treating the Coulomb term via a quartic
exciton-exciton interaction term, then coupling to light, and then
treating disorder perturbatively or not at all --- this leads to the
weakly interacting boson model, discussed in
Sec.~\ref{sec:weakly-inter-boson}.

As will be discussed further below, in the low density limit,
many features of these models are similar.
However, the different models emphasise different features: The boson
model can effectively describe the case where the dominant
interactions are exciton-exciton Coulomb interactions, while the
boson-fermion model instead has the saturation of the exciton-photon
coupling as the dominant interaction.
As such, these different models may be appropriate in different
contexts.
For example, to describe the lower polariton blue-shift, and
comparable upper polariton red-shift seen, e.g.\ in
Refs.~\cite{kasprzak06:nature,kasprzak06:thesis}, the effects of the
saturation interaction are required.
Further, the different models have been developed in different
directions, for example the effects of exciton spin, and thus the
polarisation dynamics have so far only been considered in the bosonic
model (see e.g.\ Refs.~\cite{krizhanovskii06,laussy06} and Refs.
therein).

\subsubsection{Weakly interacting boson models}
\label{sec:weakly-inter-boson}

A weakly interacting Bose gas model of polaritons can be achieved by
making an Usui transformation~\cite{usui59,hanamura77}, choosing the
bosonic operators to represent bound exciton states, and then
truncating the interaction terms at fourth order~\cite{rochat00:prb}.
This results in an effective Hamiltonian describing bosonic
excitons coupled to photon modes:
\begin{eqnarray}
  \label{eq:bosonic-model}
   H &=& \sum_\vect{k} \left[ \omega_{\vec{k}} \psi^{\dagger}_\vect{k}
     \psi_\vect{k} + \varepsilon_{\vec{k}} D^{\dagger}_\vect{k}
     D_\vect{k} + \frac{\Omega_R}{2} (D^{\dagger}_\vect{k}
     \psi_\vect{k} + \psi^{\dagger}_\vect{k} D_\vect{k}) \right]
     \nonumber\\
   &-& \frac{\Omega_R}{2\rho_{sat}}
   \sum_{\vect{k},\vect{k}^{\prime},\vect{q}} \left[
   D^{\dagger}_{\vect{k}^{\prime} -\vect{q}}
   D^{\dagger}_{\vect{k}+\vect{q}} D^{}_\vect{k}
   \psi_{\vect{k}^{\prime}} + \psi^{\dagger}_{\vect{k}^{\prime}
   -\vect{q}} D^{\dagger}_{\vect{k}+\vect{q}} D^{}_\vect{k}
   D^{}_{\vect{k}^{\prime}} \right] \nonumber\\
&+& \sum_{\vect{k},\vect{k}^{\prime},\vect{q}}
   \frac{U_{\vect{k} - \vect{k}^{\prime},\vect{q}}}{2}
   D^{\dagger}_{\vect{k}+\vect{q}}
   D^{\dagger}_{\vect{k}^{\prime}-\vect{q}} D^{}_{\vect{k}^{\prime}}
   D^{}_\vect{k},
\end{eqnarray}
Here $D^{\dagger}_{\vect{k}}$ creates a bound exciton of energy
$\varepsilon_{\vec{k}}$, $\psi^{\dagger}_\vect{k}$ creates a cavity
photon of energy $\omega_k$, and $\Omega_R$ is the effective
exciton-photon coupling strength, or Rabi splitting.
By measuring energies from the bottom of the exciton dispersion, we
may write $\varepsilon_{\vec{k}} = k^2/2M$, and expanding the photon
dispersion to quadratic order in momentum, one may approximate
$\omega_{\vec{k}} \simeq k^2/2m + \delta$, with $\delta$ the
exciton-photon detuning.
The quartic terms in Eq.~(\ref{eq:bosonic-model}) are divided into
exciton-exciton interactions, $U_{\vect{k}-\vect{k}^{\prime},\vect{q}}$, the strength of which
can also be found by calculation of the Coulomb exchange
term~\cite{ciuti98:prb,shumway01} in the Born approximation, and a
``saturation term''(second line), which decreases the
exciton-photon coupling at large exciton densities due to the
fermionic character of the excitons~\cite{rochat00:prb}.
These quartic terms arising from the Usui transformation can be seen
as an expansion of the underlying fermionic operators in powers of
bosonic operators; this expansion is controlled by the small parameter
of the number of excitons per Bohr radius.
Note that in general these terms depends also on the spin degrees of
freedom of the constituent electron and holes. For a derivation of the
dependence on spin of the Coulomb terms, see, e.g.,
Refs.~\cite{fernandez1996,bentabou2001}.

This approach takes into account the intra-exciton Coulomb term, in
forming bound excitons, and the inter-exciton Coulomb terms as
an effective quartic interaction.
The Hamiltonian in Eq.~(\ref{eq:bosonic-model}) however neglects
disorder acting on the exciton states, and as a result finds that each
exciton state couples to a single photon state, with conserved
momentum. 
However, as discussed below in Sec.~\ref{sec:boson-ferm-gener}, and in
Refs.~\cite{marchetti06:prl,marchetti06:rrs_long}, exciton disorder
will modify this picture.
Including disorder, one finds a distribution of energies, and at each
energy a distribution of exciton-photon coupling strengths.
The exciton states which have the largest coupling strength to
the low momentum photons are found to be at energies just
below the exciton dispersion edge.
Although these states are not the most localised, i.e.\ are not the
states far in the Lifshitz tail (see later
Sec.~\ref{sec:boson-ferm-gener}), they are below the band edge
and therefore are still quite strongly localised, they decay
quickly at long distances, and they result from exciton
wavefunctions concentrated around minima of the potential.
The low energy polariton modes will be formed from a superposition of
many such localised exciton states.

In addition, the saturation term in this model, which describes the
reduction of exciton-photon coupling is taken only to the lowest
order.
This is sufficient at low enough densities; however as discussed more
fully in Sec.~\ref{sec:boson-ferm-gener}, including effects of
disorder the density at which these saturation effects become
important can be much lower than the Mott density.
Thus, a quartic description of saturation may become inadequate
at modest densities, close to those already studied experimentally.
In addition, most bosonic models of polaritons further simplify
Eq.~(\ref{eq:bosonic-model}), replacing the momentum dependent
interaction $U_{\vect{k}-\vect{k}^{\prime},\vect{q}}$ with its
strength at $\vect{k}=\vect{k}^{\prime},\vect{q}=0$.
This strength is the interaction between two excitons in the same
single particle momentum eigenstate.
If exciton eigenstates are localised, it is not obvious that replacing
all exciton-exciton Coulomb interactions with an average
strength (calculated from delocalised exciton wavefunctions) is
appropriate.   
In addition, the dominant Coulomb interaction between localised, and
therefore non-overlapping, exciton states may well be due to the
direct dipole-dipole interaction, rather than exchange terms (as it is
in the clean case).
The boson-fermion model discussed Sec.~\ref{sec:boson-ferm-gener}
handles this interaction differently --- it includes strong
on-site repulsion, and neglects inter-site repulsion; this limit is
clearly also an exaggeration, and the true effects of Coulomb will be
between these two extremes.

It is worth noting parenthetically 
that a constraint on exciton density $ \sum_{\vec{k}} \langle
D_{\vec{k}}^{\dagger} D_{\vec{k}}^{\vphantom{\dagger}} \rangle < \rho_{sat}$ is required
to make the Hamiltonian in Eq.~(\ref{eq:bosonic-model}) stable.
Without such a constraint
the free energy is unbounded from below, i.e.\ for
\begin{math}
  \left|\Psi \right> = \exp\left( \lambda \psi^{\dagger}_0 + \beta
    D^{\dagger}_0 \right) \left| 0 \right>
\end{math}, the free energy $F=\left< H - \mu N \right>$
corresponding to Eq.~(\ref{eq:bosonic-model}) is
\begin{equation}
  \label{eq:bosonic-instability}
  F = (\delta - \mu)|\lambda|^2 - \mu |\beta|^2 + \Omega_R \Re(\lambda
  \beta^{\ast})\left( 1 - \frac{|\beta|^2}{\rho_{\mathrm{sat}}}
  \right) + \frac{U_0}{2} |\beta|^4.
\end{equation}
The minimum free energy can be found for real $\lambda, \beta$, and so
re-parameterising these as $\lambda = x \sin(\chi), \beta=x
\cos(\chi)$, the quartic term in Eq.~(\ref{eq:bosonic-instability})
goes like:
\begin{equation}
  \label{eq:bosonic-instability-2}
  F_4 = \frac{U_0}{4} x^4 \cos^2\chi \left[ 1 + \cos(2\chi) -
  \frac{2\Omega_R}{U_0 \rho_{sat}} \sin(2\chi) \right].
\end{equation}
For any non-vanishing $\Omega_R$, there is a value of $\chi$ for which
this is negative and so unstable.
Physically this instability is cured by restoring higher order
contributions of the saturation interaction which prevent
$\sum_{\vec{k}} \langle D_{\vec{k}}^{\dagger}
D_{\vec{k}}^{\vphantom{\dagger}} \rangle > \rho_{sat}$.
Practically the above instability can be avoided if one diagonalises
the quadratic part of Eq.~(\ref{eq:bosonic-model}), and then projects
onto the basis of lower polariton states~\cite{rochat00:prb}.
By writing:
\begin{equation}
  \label{eq:bosonic-projection}
  \left(
    \begin{array}{cc}
      \psi^{\dagger}_\vect{k} \\ D^{\dagger}_\vect{k}
    \end{array}
  \right)
  =
  \left(
    \begin{array}{rr}
      \cos\theta_\vect{k} & -\sin\theta_\vect{k} \\
      \sin\theta_\vect{k} & \cos\theta_\vect{k}
    \end{array}
  \right)
  \left(
    \begin{array}{cc}
      U^{\dagger}_\vect{k} \\ L^{\dagger}_\vect{k} 
    \end{array}
  \right)
\end{equation}
here $L^{\dagger}_\vect{k}, U^{\dagger}_\vect{k}$ create lower and upper polaritons
respectively, and $\cos\theta_\vect{k}$, $\sin\theta_\vect{k}$ are the standard
Hopfield coefficients~\cite{yamamoto,haug}.
In order to diagonalise the quadratic part of
Eq.~(\ref{eq:bosonic-model}), one must choose
\begin{equation}
  \label{eq:unitary-angle}
  \tan(2\theta_\vect{k}) = \frac{\Omega_R}{\omega_{\vec{k}} -
  \varepsilon_\vect{k}},
\end{equation}
with $\omega_{\vec{k}}$ and $\varepsilon_{\vec{k}}$ as defined
following Eq.~(\ref{eq:bosonic-model}).
Having diagonalised the quadratic part, one may project onto the
lower polariton basis for the quartic part, giving the
effective lower polariton Hamiltonian:
\begin{eqnarray}
  \label{eq:bosonic-lp}
  H_{\rm LP} &=& \sum_\vect{k} E^{\rm LP}_\vect{k}
  L^{\dagger}_\vect{k} L^{}_\vect{k} +
  \sum_{\vect{k},\vect{k}^{\prime},\vect{q}}
  V^{\mathrm{eff}}_{\vect{k},\vect{k}^{\prime},\vect{q}}
  L^{\dagger}_{\vect{k}+\vect{q}}
  L^{\dagger}_{\vect{k}^{\prime}-\vect{q}} L^{}_{\vect{k}^{\prime}}
  L^{}_\vect{k} \\
  E^{\rm LP}_\vect{k} &=& \frac{1}{2} \left[ \left(
  \omega_{\vec{k}} +
  \varepsilon_\vect{k} \right) - \sqrt{ \left(
  \omega_{\vec{k}} -
  \varepsilon_\vect{k} \right)^2 + \Omega_R^2} \right]
  \\
  V^{\mathrm{eff}}_{\vect{k},\vect{k}^{\prime},\vect{q}} &=&
  \frac{\Omega_R}{2\rho_{sat}} \cos\theta_{\vect{k}+\vect{q}}
  \cos\theta_{\vect{k}} \left[ \cos\theta_{\vect{k}^{\prime}-\vect{q}}
  \sin\theta_{\vect{k}^{\prime}} +
  \sin\theta_{\vect{k}^{\prime}-\vect{q}}
  \cos\theta_{\vect{k}^{\prime}} \right] \nonumber\\
 &+& \frac{U}{2} \cos\theta_{\vect{k}+\vect{q}} \cos\theta_{\vect{k}}
  \cos\theta_{\vect{k}^{\prime}-\vect{q}}
  \cos\theta_{\vect{k}^{\prime}}
\label{eq:effpot}
\end{eqnarray}
Note that in order for the neglect of upper polaritons to be
valid, one must be at temperatures significantly smaller than the Rabi
splitting.
This requirement of temperature can be translated to a requirement of
low densities if one is interested in phase transitions: the density
must be low enough that the Bose condensation temperature at that
density is much less than the Rabi splitting.
It can be shown~\cite{keeling05} that this latter requirement means one
should have fewer than one polariton per wavelength of light; such a
density is already exceeded  in current experiments.

The Hamiltonian (\ref{eq:bosonic-lp}) has an effective $k$
dependent interaction  strength due to the change of Hopfield
coefficient along the lower polariton branch --- i.e.\ Coulomb
interaction becomes stronger as the polariton becomes more excitonic,
and saturation interaction is strongest nearest to equal photon and
exciton components.
Preserving a $k$ dependent coupling strength requires one to think
carefully about regularisation.  
In atomic gases, the weakly interacting Bose gas model is generally
studied with a contact interaction, as is appropriate when the
scattering length is much less than the de Broglie wavelength; this is
renormalised by matching the scattering length to the experimentally
measured quantity~\cite{pitaevskii03}.
If the interaction is instead found from a microscopic theory, as
it is the case here, that microscopic theory must also describe
the regularisation of the interaction at large momentum, as a single
measured scattering length would not allow fitting of the different
momentum dependencies associated with Coulomb and saturation terms.
In practice, this means any attempt to preserve the effect of Hopfield
coefficients on the interaction must also take into account the
decrease of both Coulomb and saturation effects for large exchanged
momenta.

The limits of validity of this Hamiltonian come from several sources;
the requirement for density to be less than the Mott density $1/a_B^2$
is the easiest to understand, but also the most easily satisfied.
Neglect of the upper polariton required temperatures less than the
Rabi splitting (which translates to densities less than one polariton
per square wavelength of light); but inclusion of the upper polariton
leads to instabilities, which would require higher order terms
in the Hamiltonian to restore stability.
Thus, consideration of the phase boundary at high densities, at which
the naive estimate of the transition temperature would be comparable
to the Rabi splitting, would require a treatment beyond that
considered in this section.
Thus, in the next section we discuss an alternative model that should
be valid at these higher, yet still experimentally accessible
densities, and also takes account of the effect of disorder on the
saturation interaction.

\subsubsection{Boson-fermion, and generalised Dicke model}
\label{sec:boson-ferm-gener}

By considering first the effects of disorder acting on the excitons,
one finds that in 2D systems the effect of disorder is particularly
profound and that formally any arbitrarily small amount of disorder
leads to localisation~\cite{abrahams79,lee85}.
However, the character of the states changes significantly with
energy.
At high energies states may be described as a random superposition of
plane waves with the same modulus of momentum, and localisation
effects are weak.
At very low energies, well below the band edge, the Lifshitz tail
states~\cite{lifshitz64,halperin66,zittartz66} have a nodeless form,
localised in deep minima.
The changing nature of the exciton states with energy also changes
their oscillator strength~\cite{runge98,runge00}, and the exciton
states that couple most strongly to the long wavelength radiation
modes are those just below the band edge, for which localisation
effects are important.
As a result, those exciton states which contribute most to the
relevant (thermally populated) polariton states are effectively localised
exciton states~\cite{marchetti06:prl,marchetti06:rrs_long}.

This localisation may also be expected to modify details of the
inter-exciton Coulomb interaction term compared
to the clean picture~\cite{ciuti98:prb,rochat00:prb}.
Considering strongly localised exciton states, since exchange requires
wavefunction overlap, one expects a difference between the strength of
on-site Coulomb repulsion --- i.e.\ interaction of excitons localised
in the same potential fluctuation --- as compared to inter-site
interactions.
Taking the extreme form of this difference --- i.e.\ on-site exclusion
and neglect (or perturbative treatment) of the inter-site interaction
--- leads one to a
generalisation~\cite{marchetti06:prl,marchetti06:rrs_long} of the
Dicke model~\cite{dicke54,eastham00:ssc,eastham01}, describing
two-level systems coupled to a bosonic field:
\begin{eqnarray}
  \label{eq:dicke-model-tls}
  \hat{H} &=&
  \sum_{\alpha} 
  \varepsilon_{\alpha} S_{\alpha}^z
  + 
  \sum_{\vect{p}} 
  \omega_{\vect{p}} \psi_{\vect{p}}^\dag
  \psi_{\vect{p}} 
  +
  \frac{1}{\sqrt{A}} 
  \sum_{\alpha}
  \sum_{\vect{p}} \left(
    g_{\alpha , \vect{p}} \psi^{}_{\vect{p}} S_{\alpha}^{+}
     + \mathrm{H.c.}
  \right).
\end{eqnarray}
Here $\hat{\vec{S}}_{\alpha}$ is a spin $1/2$, representing a two-level
system, where $\left|\downarrow_{\alpha} \right>$ is the ground state --- i.e.\
no exciton on site $\alpha$ --- and $\left|\uparrow_{\alpha} \right>$ 
indicates the presence of an exciton on site $\alpha$.
Such a model has also been studied in the related context of
spontaneous superradiance~\cite{hepp73:AnnPhys,hepp73:pra,wang73}; it
was however later shown~\cite{rzazewski75} that including higher order
terms beyond the dipole approximation prevent the superradiant
transition of the vacuum state of such a model.
No such problem however occurs when one considers the system in
contact with a reservoir that fixes particle density --- the effects
discussed in Ref.~\cite{rzazewski75} apply to the stability of the
vacuum state, i.e.\ with chemical potential going to negative
infinity.

It is often convenient to represent the two-level systems as two
fermionic states so that the ground state is
$\left| \downarrow_{\alpha} \right> = a_{\alpha}^\dag \left|0\right>$, and the excitonic
state $\left| \uparrow_{\alpha} \right> = b_{\alpha}^\dag \left|0\right> =
b_{\alpha}^\dag a_{\alpha} \left|\mathrm{g.s.}\right>$. 
Imposing a constraint on total fermion occupancy, $b_{\alpha}^\dag
b_{\alpha} + a_{\alpha}^\dag a_{\alpha} = 1$, eliminates the
unphysical states $\left|0\right>$ and $a_{\alpha}^\dag b_{\alpha}^\dag
\left|0\right>$, thus giving the Hamiltonian
\begin{eqnarray}
  \label{eq:dicke-model-fermions}
  \hat{H} &=&
  \sum_{\alpha} 
  \frac{\varepsilon_{\alpha}}{2} 
  \left(b_{\alpha}^\dag b_{\alpha} + a_{\alpha} a_{\alpha}^\dag\right)
  + 
  \sum_{\vect{p}} 
  \omega_{\vect{p}} \psi_{\vect{p}}^\dag
  \psi_{\vect{p}} 
  \nonumber\\ 
  &+&
  \frac{1}{\sqrt{A}} 
  \sum_{\alpha}
  \sum_{\vect{p}} \left[
    g_{\alpha , \vect{p}} \psi^{}_{\vect{p}}   b_{\alpha}^\dag a_{\alpha}
     + \mathrm{H.c.}
  \right].
\end{eqnarray}
This formalism is easy to use, as one may show~\cite{popov88} that the
constraint preventing double occupation can be easily incorporated in
the imaginary time path integral formalism by shifting the fermionic
Matsubara frequencies according to
\begin{equation}
  \epsilon_n = (2n + 1)\pi/\beta \mapsto \epsilon_n = (2n +
3/2)\pi/\beta \;.
\end{equation}
It is important not to confuse these fermionic states (which represent
the two levels of a two-level system) with the conduction and valence band
states in Eq.~(\ref{eq:eh-hamiltonian}).
While $b^{\dagger}_{\alpha} a^{}_{\alpha}$ creates an exciton, one
should not think of $b^{\dagger}_{\alpha}$ ($a^{}_{\alpha}$) as
creating an electron (hole) --- i.e.\ one cannot write
$b^{\dagger}_{\alpha}$ ($a^{}_{\alpha}$) as a linear combination
of the electron creation operators $c^{\dagger}_{\vec{k}},
v^{\dagger}_{\vec{k}}$ ($c^{}_{\vec{k}}, v^{}_{\vec{k}}$)
 in Eq.~(\ref{eq:eh-hamiltonian}).  
If $b^{\dagger}_{\alpha}$ ($a^{}_{\alpha}$) were a linear
combination of electron (hole) creation operators, then
$b^{\dagger}_{\alpha} a^{}_{\alpha}$ would create an electron-hole
pair, but without any correlation between the position of electron and
hole --- i.e.\ without excitonic binding.
Instead, the relation between $b^{\dagger}_{\alpha} a^{}_{\alpha}$ ---
the fermionic representation of saturable excitons --- and the
underlying electrons and holes is as discussed later in
Eq.~(\ref{eq:wf-tls}) and Eq.~(\ref{eq:wf-tls-exciton}).

\begin{figure}
  \begin{center}
    \includegraphics[width=1\linewidth]{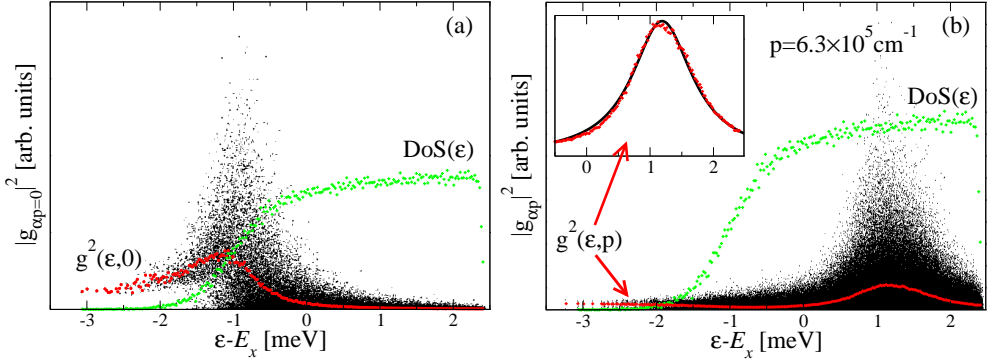}
  \end{center} 
  \caption{Distribution of energies, and energy dependence of
    exciton--light coupling strength, for excitons in the presence of
    disorder.  Black points mark the energy and coupling strength of
    individual exciton states (combined from 160 separate disorder
    realisations); green lines show density of states, red lines show
    mean square coupling strength.  Panel (a)
    [From~\cite{marchetti06:prl} Copyright (2006) by the American
    Physical Society] coupling to zero-momentum photons.  Panel (b)
    [From~\cite{marchetti06:rrs_long}.] coupling to high momentum
    photon states (much beyond relevant states).  Inset illustrates
    comparison of numerical values for the mean squared
    oscillator strength to the first Born approximation.}
  \label{fig:allg}
\end{figure}

This model naturally allows one to consider a distribution of
excitonic energies, and a distribution of excitonic oscillator
strengths for each given energy, which are set by
disorder~\cite{marchetti06:prl,marchetti06:rrs_long}.
To perform such a calculation, $\varepsilon_{\alpha}$ and
$g_{\alpha,\vect{p}}$ should be calculated by solving the
Schr\"odinger equation for the exciton centre of mass
coordinates in a random disorder potential,
\begin{equation}
  \label{eq:disorder-schrodinger}
  \left[-\frac{\nabla_{\vect{R}}^2}{2 M} + W(\vect{R})\right]
  \Phi_{\alpha} (\vect{R}) = \varepsilon_{\alpha} \Phi_{\alpha}
  (\vect{R}).
\end{equation}
and then calculating the oscillator strength from
\begin{equation}
  \label{eq:disorder-oscillator-strength}
  g_{\alpha ,\vect{p}} = e \mu_{cv} \sqrt{\frac{\omega_{\vect{p}}}{2
      \epsilon_r \epsilon_0 L_w}} \varphi_{1s} (0)
      \Phi_{\alpha,\vect{p}}.
\end{equation}
Energies and oscillator strengths calculated this way are shown in
Fig.~\ref{fig:allg}.
It is of course perfectly possible to write a theory of
disorder-localised exciton states that are bosonic modes --- and as
will be shown below, a bosonic model can be extracted from this
boson-fermion model at very low densities --- however in such a
treatment it is important to consider, as discussed above, how the
change of exciton wavefunctions modifies the Coulomb 
interaction between excitons.

A relationship between the model of this section, and that of the
previous section, may be established in the limit of low densities, by
considering a Holstein-Primakoff transformation of the Hamiltonian in
Eq.~(\ref{eq:dicke-model-tls}); i.e.\ 
\begin{equation}
  \label{eq:holstein-primakoff}
  S_{\alpha}^z =  
  D^{\dagger}_{\alpha} D^{}_{\alpha} - \frac{1}{2}, \quad
  S_{\alpha}^+ = 
  D^{\dagger}_{\alpha}
  \sqrt{1 - D^{\dagger}_{\alpha} D^{\vphantom{\dagger}}_{\alpha}}, \quad
  S_{\alpha}^- =  \left( S_{\alpha}^+ \right)^{\dagger}
\end{equation}
Then, assuming the occupation of excitons to be small (i.e.\ $\left<
D^{\dagger}_{\alpha} D^{}_{\alpha} \right> \ll 1$), one may expand
Eq.~(\ref{eq:dicke-model-tls}) to get:
\begin{eqnarray}
  \label{eq:dicke-model-holstein-primakoff}
  \hat{H} &=&
  \sum_{\alpha} 
  \varepsilon_{\alpha} D^{\dagger}_{\alpha} D^{}_{\alpha}
  + 
  \sum_{\vect{p}} 
  \omega_{\vect{p}} \psi_{\vect{p}}^\dag
  \psi_{\vect{p}} 
  \nonumber\\ 
  &+&
  \frac{1}{\sqrt{A}} 
  \sum_{\alpha}
  \sum_{\vect{p}} \left(
    g_{\alpha , \vect{p}} \psi^{}_{\vect{p}}   D^{\dagger}_{\alpha}
    \left(
      1 - \frac{1}{2}D^{\dagger}_{\alpha} D^{\vphantom{\dagger}}_{\alpha}
    \right)
    + \mathrm{H.c.}
  \right).
\end{eqnarray}
Comparing this to Eq.~(\ref{eq:bosonic-model}) shows that a bosonic
model derived in this way has certain differences to the standard
bosonic model;  it obviously neglects the inter-exciton Coulomb
term, as this was neglected in Eq.~(\ref{eq:dicke-model-tls}),
and the exciton energies are set by localised states in a disorder
potential $\varepsilon_{\alpha}$, rather than $\vect{k}^2/2 M$.
Less obviously, but more importantly, the saturation interaction term
is significantly stronger than would be suggested by
Eq.~(\ref{eq:bosonic-model});  in that case, the mean-field energy shift
at polariton density $n$ is of the order of
\begin{equation}
  \label{eq:clean-blueshift}
    \delta E^{\rm LP}_{\mathrm{sat}} \sim \Omega_R n a_{\ex}^2.
\end{equation}
In contrast, the term in Eq.~(\ref{eq:dicke-model-holstein-primakoff}) is
of the order of
\begin{equation}
  \label{eq:dirty-blueshift}
  \delta E^{\rm LP}_{\mathrm{sat}} \sim \Omega_R n \xi_d^2\; 
\end{equation}
where $\xi_d \sim (M W_{\rho})^{-1/2}$ is a characteristic length
scale of the disorder potential.
This energy shift is important as it relates the observed lower
polariton blue-shift to the polariton density, and so is important
in the interpretation of experiments.
This result is valid at low temperatures; at higher temperatures one
can show~\cite{marchetti06:rrs_long}, that $\xi_d$ should be replaced
by $\xi_T \sim (M k_B T)^{-1/2}$.
The appearance of this temperature-dependent length scale would not
arise from a model that included only bosonic lower polaritons.
Comparison of the equilibrium transition temperatures of the
two models is discussed later, in Sec.~\ref{sec:t_c-phase-boundary}.

\subsubsection{Comparison of models}
\label{sec:comparison-models}

As is clear from the above discussion, the Bose-Fermi
model in some sense encompasses a bosonic model.
However, its derivation led naturally to the inclusion of saturation
interaction, but as yet no generalisation including long-range Coulomb
interaction has been studied.
The question of comparing the models is therefore not so much whether
one model is right or wrong, but whether interaction effects beyond a
quartic boson-boson interaction are important, and so whether a
description like that of Eq.~(\ref{eq:dicke-model-tls}) is necessary.
At low enough densities and temperatures (i.e.\ temperatures a small
fraction of the Rabi splitting) it is clear such a description is not
necessary.
However, the definition of ``low enough'' that is derived from
studying when the (equilibrium) phase boundary of
Eq.~(\ref{eq:dicke-model-tls}) is reproduced by a bosonic theory
suggests that low enough means exciton separation of the order of the
wavelength of
light~\cite{keeling04:polariton,littlewood04:jpcm,keeling05}, rather
than, e.g.\ the exciton Bohr radius; and temperatures of the order of
tenths of the Rabi splitting.

As an alternative way to resolve the question of which approximate
Hamiltonian, Eq.~(\ref{eq:dicke-model-tls}) or
Eq.~(\ref{eq:bosonic-model}), is most appropriate for a given physical
system, one can propose the following clear, but technically
challenging approach.
From both Hamiltonians, one can construct an approximate ground state,
which can then be rewritten in terms of electrons, holes and photons.
In both cases, we consider generalisations of the coherent state,
which for a simple structureless boson field $L^{\dagger}$ would
be written $\exp(\lambda L^{\dagger}) \left| 0 \right>$.
This leads to two different trial wavefunctions for the
electron-hole-photon system.
While this is not a simple exercise --- and would in fact require
extensive numerical computation --- it is a useful gedanken comparison
to highlight the distinctions.
Let us consider first the trial wavefunction appropriate to the Hamiltonian
of Eq.~(\ref{eq:bosonic-lp}).
Taking $\left|0\right>$ as the filled valence band,  we have
\begin{equation}
  \label{eq:wf-boson}
 \left| \Psi_{Bose} \right>
  =  
  e^{  \lambda L^{\dagger}_0 } \left| 0
  \right>
  ; \quad
  L^{\dagger}_0
  =
  \cos(\xi_0) \psi ^{\dagger}_0 +
  \sin(\xi_0) \sum_\vect{q} \tilde\varphi(\vect{q})
  c^{\dagger}_{\vect{q}} v^{}_{-\vect{q}} .
\end{equation}
At low densities this wavefunction has a simple interpretation;
$\tilde{\varphi}(\vect{q})$ is the bound exciton wavefunction,
and the $\xi_0$ controls the exciton and photon fractions of the
lower polariton;  i.e.\ the term in brackets is the lower polariton
creation operator, and this is a coherent state of lower polaritons.
Note however that $(c^{\dagger}_{\vec{q}} v_{-\vec{q}})^2=0$, as
$c^{\dagger}_{\vec{q}},v_{-\vec{q}}$ are fermionic operators, thus
this wavefunction can be also written as:
\begin{equation}
  \label{eq:wf-boson-rewrite}
  \left| \Psi_{Bose} \right>
  =
  \exp \left( 
    \lambda  
    \cos(\xi_0) \psi ^{\dagger}_0 
  \right)
  \prod_{\vec{q}}
  \left(
    1 
    +
    \lambda \sin(\xi_0)
    \tilde\varphi(\vect{q}) c^{\dagger}_{\vect{q}} v^{}_{-\vect{q}}
  \right)
  \left|
    0
  \right>.
\end{equation}
Thus, if $\tilde{\varphi}(\vect{q})$ has a step-like form, this can
also describe a BCS-like state~\cite{nozieres85,nozieres}.
More generally, the parameters $\lambda, \xi_0$ and the function
$\tilde{\varphi}(\vect{q})$ can be taken as variational parameters,
and used to minimise the energy.

Starting instead from the Hamiltonian of
Eq.~(\ref{eq:dicke-model-tls}) one is instead led to write:
\begin{eqnarray}
  \label{eq:wf-tls}
  \left| \Psi_{TLS} \right>
  &=&
  e^{ \lambda  \psi ^{\dagger}_0}
  \prod_{\alpha} 
  \left(
    \cos(\theta_{\alpha})
    +
    \sin(\theta_{\alpha})  D^{\dagger}_{\alpha}
  \right)
  \left| 0 \right>_{TLS}
  \\
  \label{eq:wf-tls-exciton}
  D^{\dagger}_{\alpha} 
  &=&
  \sum_{\vect{k},\vect{q}} \tilde{\Phi}_{\alpha}(\vect{k}) \tilde\varphi(\vect{q}) 
  c^{\dagger}_{{m_e} \vect{k}/({m_e + m_h}) + \vect{q}} 
  v^{}_{{m_h} \vect{k}/({m_e + m_h}) - \vect{q}}
\end{eqnarray}
where we have now introduced $\tilde{\Phi}_{\alpha}(\vect{k})$ as the
localised centre of mass wavefunction.
Note that the operator, $D^{\dagger}_{\alpha}$, describing a localised
exciton does not square to zero.
It is thus not possible to rewrite the BCS-like product in
Eq.~(\ref{eq:wf-tls}) as an exponential; there is a qualitative
difference between the states in Eq.~(\ref{eq:wf-boson-rewrite}) and
Eq.~(\ref{eq:wf-tls}).
Although $(D^{\dagger}_{\alpha})^2 \ne 0$, the product in
Eq.~(\ref{eq:wf-tls}) only allows each operator $D^{\dagger}_{\alpha}$
to occur at most once, so for a given single-particle state labelled
by $\alpha$, only zero or one excitons may occupy it, and thus prevents
multiple occupation.
By including the disorder-localised centre of mass wavefunctions,
Eq.~(\ref{eq:wf-tls}) describes single occupation of a set of
localised exciton wavefunctions, while in comparison,
Eq.~(\ref{eq:wf-boson}) describes only the single, lowest energy,
delocalised exciton mode.
As above, we may take the parameters $\lambda, \theta_{\alpha}$ and
the functions
$\tilde{\Phi}_{\alpha}(\vect{k}),\tilde{\varphi}(\vect{q})$ as
variational.

Unfortunately, direct evaluation of the expectation of
Eq.~(\ref{eq:eh-photon-hamiltonian}) with these trial wavefunctions is
challenging.
At low enough densities, no multiple occupation occurs, so in
this limit Eq.~(\ref{eq:wf-boson}) and Eq.~(\ref{eq:wf-tls})
become comparable:
Expanding Eq.~(\ref{eq:wf-tls}) for small $\theta_{\alpha}$,
the terms in the product can be rewritten approximately
as:
\begin{equation}
  \label{eq:wf-tls-bosonic}
  \prod_{\alpha}\left(
    1 + \theta_{\alpha} D_{\alpha}^{\dagger}
  \right)
  + \mathcal{O}(\theta_{\alpha}^2)
  \simeq
  \exp\left[
    \sum_{\alpha} \theta_{\alpha}  D_{\alpha}^{\dagger}
  \right] 
  + \mathcal{O}(\theta_{\alpha}^2).
\end{equation}
This would be equivalent to Eq.~(\ref{eq:wf-boson}) except that
Eq.~(\ref{eq:wf-boson}) macroscopically occupies the $\vec{k}=0$ exciton
state, whereas Eq.~(\ref{eq:wf-tls-bosonic}) occupies a collection of
disorder-localised states.
Although not identical, a superposition of many localised states
distributed across the sample can (at low enough densities) behave 
similarly to the translationally invariant $\vec{k}=0$ state.
Thus depending on the relative importance of disorder localisation,
and on the difference of Coulomb interaction between different
single-particle exciton states vs interaction for multiple occupation
of the same single-particle state, one may find which of
Eq.~(\ref{eq:wf-boson}) or Eq.~(\ref{eq:wf-tls}) has lower energy.

Furthermore, both of the above wavefunctions are mean-field approximations
of the ground state, and in both cases, energy could be lowered by
constructing the Nozi\`eres-Bogoliubov state.
To discuss this, let us consider the simpler case of
structureless bosons, $L^{\dagger}_\vect{k}$.
One can then understand this state in two ways, either as a
variational ansatz, as in Ref.~\cite{nozieres82}:
\begin{equation}
  \label{eq:nozierres-state}
  \left| \Lambda \right> = \exp\left( \lambda L^{\dagger}_0 +
  \sum_\vect{k} \lambda_\vect{k} L^{\dagger}_\vect{k}
  L^{\dagger}_{-\vect{k}} \right) \left| 0 \right>,
\end{equation}
and then find $\lambda, \lambda_\vect{k}$ by minimisation.
Alternatively, the same state can be described if one considers
fluctuation corrections to the mean-field theory.
As is well known, in the presence of a condensate, the quasi-particles
are the Bogoliubov modes~\cite{pitaevskii03},  i.e.\
\begin{equation}
  \label{eq:bogoliubons}
  B^{\dagger}_\vect{k} = \cosh(\phi_\vect{k}) L^{\dagger}_\vect{k} +
  \sinh(\phi_\vect{k}) L^{}_{-\vect{k}},
\end{equation}
with $\phi_\vect{k}$ the Bogoliubov rotation angle.
Thus, given the Bogoliubov spectrum, the lowest energy state is the
Bogoliubov vacuum, $\left| \Omega_{\mathrm{Bog}} \right>$, defined such
that it is annihilated by all $B^{}_\vect{k}$, i.e.\
\begin{equation}
  \label{eq:bog-vacuum-eqn}
  \left[ \cosh(\phi_\vect{k}) L^{}_\vect{k} + \sinh(\phi_\vect{k})
  L^{\dagger}_{-\vect{k}} \right] \left| \Omega_{\mathrm{Bog}} \right>
  = 0 \quad \forall \vect{k}\; ,
\end{equation}
which is clearly solved by:
\begin{equation}
  \label{eq:bog-vacuum-soln}
  \left| \Omega_{\mathrm{Bog}} \right> = \exp\left( - \sum_\vect{k}
  \tanh(\phi_\vect{k}) L^{\dagger}_\vect{k} L^{\dagger}_{-\vect{k}}
  \right) \left|0\right>
\end{equation}
Two comments are in order about the significance of this state;
firstly, the physical reason this state is of lower energy is the
quartic interaction, in particular terms like $L^{\dagger}_\vect{k}
L^{\dagger}_{-\vect{k}} L_0 L_0 + H.c.$, which favour states which are
not eigenstates of the number of $\vect{k}=0$ particles.
Secondly, even when projected to an overall number state, one may
retain features of this state, by writing a superposition of terms
with different division of the number of particles between the
condensate mode and other states.

\subsection{Theoretical treatments --- effects of the environment}
\label{sec:theor-treatm-effects}

Having discussed various models of the polariton system,
we now turn to how these models, and the effects of the
environment, may be treated.
We first briefly outline the thermal equilibrium case,
and compare mean-field theories of the two models
discussed above.
We then discuss some of the various approaches that one may use to
describe the effects of the environment, focusing mainly on
non-thermal steady states.
Finally, we try to separate and clarify the concepts of coherence,
condensation, superfluidity and lasing, which while often related,
need not necessarily occur together.

\subsubsection{Thermal equilibrium}
\label{sec:therm-equil-vs}

The simplest approximation for the environment is to consider the
system in thermal and chemical equilibrium with a bath.
While it is clear that the current experiments involve substantial
pumping and decay, which will be discussed next, there are compelling
reasons to deal with the equilibrium case.
Firstly, the properties of a given model in the equilibrium case are
instructive when considering the range of behaviour it can show; while
the equilibrium properties of weakly interacting dilute Bose gas are
well studied~\cite{popov,zzz_bec,pitaevskii03}, the properties of
models like Eq.~(\ref{eq:dicke-model-tls}), with distributions of
oscillator strengths and
energies~\cite{marchetti06:prl,marchetti06:rrs_long} are less
known.
Even within the weakly interacting Bose gas picture, interesting
features can arise from considering non-quadratic
dispersion~\cite{kavokin03:pla,malpuech03,keeling06}, or the effects
of anisotropic spin interactions~\cite{laussy06,rubo06}.
The second reason is that with improvements in the quality of mirrors,
and refinement to the design of microcavities and the conditions of
pumping, experiments have been able to increase the thermalisation
rate to be comparable to or faster than polariton decay
rates~\cite{kasprzak06:nature,deng06:eqbm}, and so for these, or
future, experiments, the correct description may become increasingly
close to equilibrium.

The treatment of both Eq.~(\ref{eq:bosonic-model}) and
Eq.~(\ref{eq:dicke-model-tls}) in equilibrium can be put in a similar
form by considering their saddle point, or minimum action equations.
Formally, these can be derived by writing the imaginary
time path integral for the partition function~\cite{nagaosa1},
and then considering the configurations that minimise the
imaginary time action.
Thus, for the bosonic case within the effective lower polariton
model, Eq.~(\ref{eq:bosonic-lp}), the saddle point solutions satisfy a
Gross-Pitaevskii equation:
\begin{equation}
  \label{eq:gpe}
  \left[ -i \partial_t + E^{\rm LP}_{\vect{k}=0} - \frac{{E^{\rm
  LP}_{\vect{k}=0}}^{\prime\prime} \nabla^2}{2} +
  \mathcal{O}(\nabla^4) + V^{\mathrm{eff}}_{0,0,0} |L_0|^2 \right] L_0
  \simeq 0
\end{equation}
Here, as the dispersion $E^{\rm LP}_\vect{k}$ is not quadratic, we have
expanded it to quadratic order to find the coefficient of $\nabla^2$.
Note that by considering solutions of the form $L_0(t) = L_0 e^{-i\mu
  t}$, one can introduce the chemical potential, and thus recover the
expected static Gross-Pitaevskii equation.

For the fermionic model, more care is required;  since there has been no
projection onto lower polaritons, the saddle point condition leads to
 coupled equation for the photon field and two-level systems.
Using the spin notation of Eq.~(\ref{eq:dicke-model-tls}) one has:
\begin{eqnarray}
  \label{eq:coupled-gpe-fermi}
  \left[ -i \partial_t + \omega_0  - \frac{\nabla^2}{2 m} 
 \right] \psi_0 = \sum_{\alpha} \frac{g_{\alpha,0}}{\sqrt{A}} S_{\alpha}^-
 \delta(\vec{r}-\vec{r}_{\alpha})
 ; 
 \\
 \partial_t \vec{S}_{\alpha} = - \vec{B}_{\alpha} \times \vec{S}_{\alpha},
 \qquad
 \vec{B}_{\alpha} = \left( 
   \begin{array}{c}
     g_{\alpha,0} (\psi_0 + \psi^{\dagger}_0) \\
     g_{\alpha,0} i(\psi_0 - \psi^{\dagger}_0) \\
      \varepsilon_\alpha
   \end{array}
 \right).
\end{eqnarray}
Here $\vec{r}_{\alpha}$ is the localisation site of the two-level
system $\vec{S}_{\alpha}$.
In the case where the only variation is $\psi_0(t) = \psi_0 e^{-i\mu
  t}$ and the polarisation has the same time variation, one can eliminate
the time variation by a gauge transformation.

The sum over exciton energy levels can also be simplified if
one makes two assumptions: firstly that the excitations
are occupied according
to a thermal distribution, and secondly that we can average over many
realisations of excitonic disorder.
This second assumption, that $\psi_0(\vec{r})$ varies slowly compared
to the distance between excitons, or equivalently that the
photon couples to many localised exciton modes, allows one to replace
the sum over exciton energy levels with a sum over the
statistical distribution of energies and the excitonic coupling
strengths.
This then yields:
\begin{equation}
  \label{eq:gpe-fermi}
  \left[ \omega_0  - \mu- \frac{\nabla^2}{2 m} +
  \mathcal{O}(\nabla^4) - 
  \sum_{\alpha} 
  \frac{g_{\alpha,0}^2}{A}
  \frac{\tanh(\beta E_{\alpha})}{2 E_{\alpha}} \right] \psi_0 \simeq 0 \; ,
\end{equation}
where the energy $E_{\alpha}^2 = (\varepsilon_{\alpha} - \mu)^2 +
g_{\alpha,0}^2 |\psi_0(\vec{r})|^2$ depends on the local value of the
slowly varying $\psi_0(\vec{r})$.
Note that in this way the exciton disorder does not lead to spatial
inhomogeneity of the polariton condensate, and so in the absence of
photonic disorder one would expect polariton condensation in the
$\vec{k}=0$ mode.
This has a clear similarity to Eq.~(\ref{eq:gpe}), but in this case,
the non-linear interaction term is more complicated than it was in the
bosonic case, $V^{\mathrm{eff}}_{0,0,0} |L_0|^2$, and the
polariton-polariton interaction is due to the nonlinearity of the
susceptibility arising from the saturable nature of the excitons.
For a uniform and static condensed solution ($\nabla^2 \psi_0 = 0 =
\partial_t \psi_0$), the  Gross-Pitaevskii equation
(\ref{eq:gpe-fermi}) is also analogous to the gap equation
(self-consistency condition) of the BCS theory~\cite{nagaosa1}.

Despite their similarity, there is an important distinction between
Eq.~(\ref{eq:gpe}) and Eq.~(\ref{eq:gpe-fermi}); Eq.~(\ref{eq:gpe}) is
temperature independent, while the nonlinear susceptibility in
Eq.~(\ref{eq:gpe-fermi}) decreases at high temperature, and is
eventually incapable of supporting condensation.
Thus one can crudely say that Eq.~(\ref{eq:gpe}) can support
mean-field condensation at any temperature, and
fluctuations~\cite{popov} must be considered to find a transition
temperature.
For the bosonic model, going beyond mean-field theory, one can also
produce a temperature dependent equation by using a
Hartree-Fock wavefunction, and thus including the effect of
interactions between the condensate and non-condensed particles (see
e.g.\ Ref.~\cite{pitaevskii03} for further details).
In distinction Eq.~(\ref{eq:gpe-fermi}) contains a finite mean-field
transition temperature, and so fluctuations are only important when
they significantly decrease this transition
temperature~\cite{keeling04:polariton,keeling05}.
Thus, including fluctuations one finds a crossover from a fluctuation
dominated phase boundary at low densities, to a phase boundary
that is well described by mean-field theory in the high density limit,
where long-range interactions dominate.
The temperature dependence that appears in Eq.~(\ref{eq:gpe-fermi})
can also be understood by noting that it was necessary to integrate
out the excitonic degrees of freedom in
Eq.~(\ref{eq:coupled-gpe-fermi}) to produce an effective action for a
single photon field.
Thus, the saddle point density for the Bose-Fermi model contains
both the condensate, and a contribution from incoherent excitons.

By considering fluctuation corrections to the saddle point density,
one may in three dimensions, and at low densities, recover the
non-interacting transition temperature of a weakly interacting Bose
gas.
i.e.\ considering the transition from the normal side, a mean-field
condensate density appears when $\mu\to 0$, and so the critical
temperature is given by $\rho_{\mathrm{total}} =
\rho_{\mathrm{fluct}}(T_c,\mu=0)$.
However, for a two-dimensional system, when the transition is of the
Berezhinskii-Kosterlitz-Thouless (BKT)
class~\cite{nelson77,kosterlitz73,fisher88}, it is necessary to
consider fluctuations in the presence of a quasi-condensate, as the
BKT transition, where free vortices proliferate, occurs below the
mean-field transition temperature.

In order to calculate the total density in the presence of a
condensate, it is important to note that the fluctuation corrections
can deplete the condensate population, as well as increase the
population of other modes~\cite{keeling05}.
The condensate density that comes from a mean-field calculation
[i.e.\ from the uniform static solutions to Eq.~(\ref{eq:gpe})]
is $\rho = |L_0|^2 = \mu/V^{\mathrm{eff}}$.
One compact way of finding how fluctuations deplete the condensate
density is by using the Hugenholtz-Pines relation as discussed in
e.g.\ Ref.~\cite{popov}.
Let us briefly summarise here how this argument shows that the
condensate density is smaller than the mean-field estimate.
To discuss this one must introduce the self energy of the condensate
$\Sigma$.
If we define the matrix of Green's functions:
\begin{equation}
  \label{eq:define-gf}
  \mathcal{G}(\omega,\vec{k}) = 
  \int dt e^{-i\omega t}
  \left< \left(
      \begin{array}{cc}
        L^{\dagger}_{\vec{k}}(t) L^{}_{\vec{k}}(0) 
        &
        L^{}_{-\vec{k}}(t) L^{}_{\vec{k}}(0) 
        \\
        L^{\dagger}_{\vec{k}}(t) L^{\dagger}_{-\vec{k}}(0) 
        &
        L^{}_{-\vec{k}}(t) L^{\dagger}_{-\vec{k}}(0) 
      \end{array}
    \right) \right>,
\end{equation}
and introduce $\mathcal{G}_0(\omega,\vec{k})$ as the free Green's
function (i.e.\ in the absence of interactions), then the matrix of
self energies is defined by $\Sigma(\omega,k) =
\mathcal{G}^{-1}_0(\omega,\vec{k}) -
\mathcal{G}^{-1}(\omega,\vec{k})$.
The Hugenholtz-Pines relation is the condition required of this self
energy in order that there might be a gapless mode, as one expects for
a Bose-condensed system.
The condition can be written as $\Sigma_{11} - \Sigma_{12} = \mu$.
By writing the self energies in terms of the densities of condensate
and normal state particles, one may use this identity to write the
condensate density in terms of $\mu$ and the normal state density.
At leading order in interaction strength, it can be shown~\cite{popov}
that the self energy is given by:
\begin{equation}
  \label{eq:hugenholz-pines}
  \Sigma_{11} = 2 V^{\mathrm{eff}} (\rho_0 + \rho_1), \quad
  \Sigma_{12} = V^{\mathrm{eff}}(\rho_0 - \tilde{\rho}_1).
\end{equation}
In this expression $\rho_0$ is the condensate density, $\rho_1$ is the
density of particles in other states, and $\tilde{\rho}_1$ is the
anomalous density, $\sum_\vect{k} \langle L^{\dagger}_\vect{k}
  L^{\dagger}_{-\vect{k}}\rangle$.
One thus finds: $\rho_0 = (\mu/V^{\mathrm{eff}}) - (2\rho_1 + \tilde{\rho}_1)$, i.e.\
fluctuations reduce the condensate density below its mean-field
value.

\begin{figure}
  \begin{center}
    \includegraphics[width=0.7\linewidth]{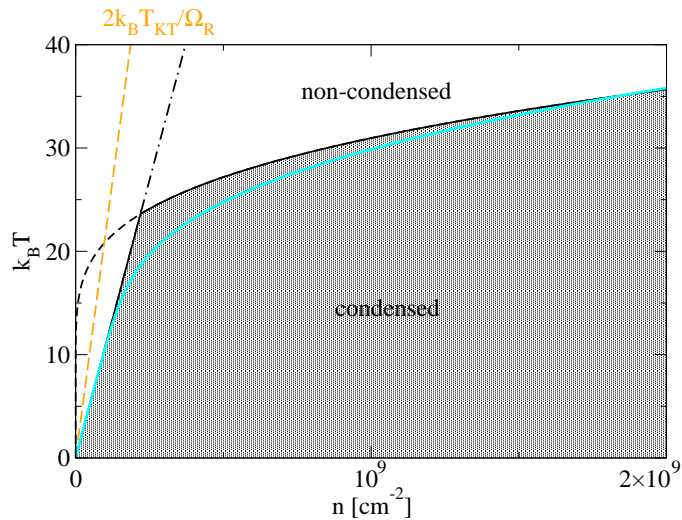}
  \end{center} 
  \caption{ Phase boundaries for an equilibrium polariton condensate.
    Solid and dashed lines mark the mean-field phase boundary of the
    boson-fermion model of
    Refs.~\cite{marchetti06:prl,marchetti06:rrs_long} (dashed line
    indicates the region in which this boundary is strongly modified
    by fluctuation corrections).  Blue lines indicate the BKT
    transition temperature for thermal population of the lower
    polariton branch, with the modified Landau criterion as discussed
    in Ref.~\cite{keeling06}. The interaction strength used for this
    curve is that which arises from the bosonic approximation to the
    boson-fermion model --- i.e.\ includes the effect of disorder in
    the saturation interaction.  For comparison, orange dashed line
    marks the phase boundary for the na\"ive estimate (i.e.\
    neglecting condensate depletion due to density fluctuations) BKT
    transition temperature for a quadratic dispersion with effective
    polariton mass.  Plotted for $\Omega_R=26 \mathrm{meV}$
    exciton--photon detuning $\delta=\omega_0 - \varepsilon^{\ast}=5.4
    \mathrm{meV}$, photon mass $m_{ph} = 2.59 \times 10^{-5} m_e$.}
  \label{fig:phased}
\end{figure}

To then extract the BKT temperature, one needs to find the condition
for free vortices to proliferate~\cite{nelson83,fisher88,stoof93}.
As described in those works, this requires one to know the fugacity of
a vortex, and the effective vortex-vortex interaction strength, both
of which depend on the superfluid stiffness $\rho_s$, which may be
found from the difference between transverse and longitudinal current
response functions (see Eq.~(\ref{eq:superfluidity})).
The result is a transition which occurs at $k_B T_{\mathrm{BKT}} =
(2/\pi) (\rho_s/m)$.  
In the case of bosons with quadratic dispersion, and in the limit of
weak quartic interaction strength $V^{\mathrm{eff}}$, i.e.\ $m
V^{\mathrm{eff}} \ll 1$, one may extract an asymptotic relation
between the superfluid density and the total density, giving $k_B
T_{\mathrm{BKT}} = [2\pi/\ln(B/m V^{\mathrm{eff}})](\rho/m)$, where
quantum Monte Carlo calculations~\cite{prokofev01} give $B=380\pm3$.
The phase boundary calculated according to the fermionic model, i.e.\
using Eq.~(\ref{eq:gpe-fermi}), and the boundary for the BKT
transition in a bosonic model following
Refs.~\cite{kavokin03:pla,malpuech03,keeling06}, but with the
effective inclusion of disorder, are shown in Fig.~\ref{fig:phased}.
These boundaries are discussed further in
Sec.~\ref{sec:t_c-phase-boundary}.
%

\subsubsection{Pumping, decay, and non-equilibrium treatments}
\label{sec:dynam-boltzm-equat}

A more realistic discussion of the experimental environment must
consider that polaritons may escape, and so continuous pumping is
required to produce a steady state.
In addition, if one is to describe pulsed experiments, or the
transient behaviour after the pump  is switched on, a dynamical
approach is required to describe the time dependence of population~\cite{mieck02,cao04,doan05:prb,rubo06}.
Considering for the moment steady state situations --- i.e.\  c.w.\
(continuous wave) pumping --- one may highlight two important features
of the difference between the pumped, decaying system and thermal
equilibrium.
The first is that the distribution function; i.e.\ the population of
each energy level, may be far from thermal, and set instead by the
balance of pumping, decay, and thermalisation
rates;~\cite{tassone97:prb,tassone99,porras02,doan05:prb,szymanska06:long}.
The second class of effects is that incoherent pumping and decay
introduce dephasing, and can change the excitation spectrum of the
system, the additional inclusion of these effects are discussed in
Refs.~\cite{szymanska06:prl,szymanska06:long}, (see also the
discussion in Sec.~\ref{sec:phase-degree-freedom}).
There are a wide variety of approaches that may be applied to study
one or either of these features; in the following we discuss briefly
how some of these various approaches are related, and what limitations
they may have.
For a more general discussion see e.g.\
Refs.~\cite{kadanoff62,gardiner,kamenev05}.

In order to describe the properties of the pumped, decaying system, 
one requires a method to calculate various correlation functions.
Given an expression for single particle correlation functions, one may
then find many properties of interest, for example the occupation of
modes, the luminescence and absorption spectra, and the first-order
coherence properties.
The most general information about one particle correlations can be
written in terms of the two correlation functions
$\mathcal{G}^{<}(t,\vect{r}) = \left< \psi^{\dagger}(t,\vect{r})
\psi^{}(0,0)\right>$, $\mathcal{G}^{>} = \left< \psi^{}(t,\vect{r})
\psi^{\dagger}(0,0)\right>$, which with $\psi$ describing the photon
field correspond directly to luminescence and absorption
probabilities.
These encode information both about the form of the spectrum and thus
the density of states, and also about the population of those states.
For example, the density of states is given by
$\Im[\mathcal{G}^{R}(\omega+i0,\vect{k})]$, where the retarded Green's
function can be written as: $\mathcal{G}^R(t,\vect{r}) =
\Theta(t)[\mathcal{G}^>(t,\vect{r}) - \mathcal{G}^<(t,\vect{r})]$.
In equilibrium, these two Greens functions can be related in terms of
the thermal distribution function, but out of equilibrium no such
simplification is possible.

Let us now discuss different methods to calculate these Green's
functions.
The first method is to find and solve the operator equations of motion
for $\psi,\psi^{\dagger}$, and thus to evaluate the correlation
functions directly.
In order to describe pumping and decay, one considers coupling the
system to baths, which either pump particles and energy into the
system, or provide modes into which particles may decay.
These baths are assumed to be large, so their properties (e.g.
distribution functions) are fixed, and not affected by the system.
Since the bath and system are coupled, the equations of motion
for the system operators will also include bath operators.
If one considers the initial state of the bath to be drawn from some
fixed (e.g.\ thermal) distribution, then the expectation of bath
operators will be random quantities, with statistical properties set
by the bath's distribution~\cite{mieck02,gardiner}; thus such coupling
to baths introduces noise, giving quantum Langevin
equations~\cite{gardiner}.
The second method is to write equations of motion for Green's
functions, which will now involve correlation functions of
bath operators, which again can be found if one assumes a fixed
distribution for the baths.
These coupled equations are the content of the Keldysh
formalism~\cite{keldysh65}, in which it turns out to be simpler to
write equations for various linear combinations of the Green's
functions, allowing one to combine the Green's functions in a compact
matrix notation (see e.g.~\cite{kadanoff62}).
The equations for the Green's functions can also be derived 
in a path integral formulation~\cite{kamenev05}.
The path integral formalism allows one to make a close connection with
the methods discussed in Sec.~\ref{sec:therm-equil-vs}, and can
describe changes to both the occupation and to the spectrum induced by
pumping and decay~\cite{szymanska06:prl,szymanska06:long}.
As such, it allows one to introduce a complex self-consistency
condition (i.e.\ a complex equivalent of the Gross-Pitaevskii
equation); this interpolates between the laser, in which
self-consistency requires balancing of pumping and decay, and the
equilibrium condensate, where self-consistency instead relates real
self-energy shifts.
This point will be discussed in more detail in
Sec.~\ref{sec:phase-degree-freedom}.

The above two approaches allow one to find self-consistently the
population and the spectrum in the presence of pumping and decay.
In certain cases --- for example in the normal state, with weak
pumping and decay --- the changes to the spectrum may be small, and
one is interested primarily in changes to the population.
By assuming a known form for the retarded Green's function (i.e.\
for the spectrum of the system) it is possible to extract equations
for the population from the Green's function
formalism (see Ref.~\cite[Chapter 9]{kadanoff62} for details).
These will lead to the Boltzmann equation~\cite{Lifshitz:Phys_Kin},
which may also be derived phenomenologically, by considering
various rates of transfer between different energies.
Thus, by neglecting the ``kinetic'' effects, that change the spectrum,
but retaining ``dynamic'' effects, that change the populations, one
may investigate how pumping and decay change the occupation of the
spectrum~\cite{tassone97:prb,tassone99,porras02,cao04,doan05:prb}.
Even without pumping and decay, the spectrum changes due to
interactions in the condensed state~\cite{pitaevskii03}; and further
calculations retaining ``kinetic'' effects suggest the pumping and
decay further modify the spectrum (see
Refs.~\cite{szymanska06:prl,szymanska06:long} and also
Ref.~\cite{wouters05} in a different context).
Because changes to the spectrum modify the density of states, they can
be expected to in turn affect the population dynamics; it is
therefore not clear how valid it is to consider population dynamics in
the condensed system without self-consistent treatment of the changes
to the spectrum.

A different approach to treating the coupling to baths is to consider
the density matrix, which allows calculation of any single-time
expectation of operators, as well as certain multi-time correlations,
discussed further below.
Calculating the full $N$-body density matrix allows calculation of
correlation functions of arbitrary numbers of particles, rather 
than just the single particle correlations discussed above.
Density matrix methods, and Green's functions methods can be related
as the single-particle density matrix (i.e.\ tracing over coordinates
of all but one particle) is equivalent to the equal time part of the
one particle Green's function.

To numerically calculate evolution of the density matrix, due both to
the system Hamiltonian, and to  the effect of coupling to baths, one can
choose an appropriate basis and write the density matrix in terms of a
distribution function over this basis, i.e.\ $\hat{\rho} = \sum
\left|a\right> \left< b \right| P(a,b)$.
The time evolution of the density matrix then corresponds to
the time evolution of this distribution function.
Under certain conditions~\cite{gardiner,steel98}, it is possible to
interpret the evolution of this distribution as describing evolution
of a quasi-probability distribution.
In such a case, the equation of motion for $P(a,b)$ is a Fokker-Planck
equation, and can be rewritten in terms of Langevin equations for
classical variables.
If one chooses to resolve the density matrix onto a basis of coherent
states, a variety of ways of doing this exist, among which we mention
two important choices: the positive P distribution, and the Wigner
distribution (See~\cite{gardiner} for other possible choices, and
further details).
The positive P distribution formally has the desired properties, and
gives the desired Langevin equations, but is numerically unstable when
applied to the kind of problem discussed here~\cite{steel98}.
The Wigner distribution, although not technically matching the above
requirements, can have its equation of motion approximated --- the
truncated Wigner representation~\cite{steel98,carusotto05} --- which
allows one to find appropriate Langevin equations, and is numerically
stable.

The Wigner distribution allows one to find the expectation of
symmetrised products of operators; at equal times it is trivial to
extend this to find general products of operators, by making use of
equal time commutation relations.
It is also possible to find multi-time correlation functions from such
an approach, however since the unequal time commutation relations are
not a-priori known, one cannot generally find expectations of other
orders; in this sense the truncated Wigner approach does not allow
$\mathcal{G}^{<}$ and $\mathcal{G}^{>}$, but only the symmetrised
combination $\mathcal{G}^{<} + \mathcal{G}^{>}$, and hence cannot
separate the density of states from its occupation.

\subsubsection{Lasing, condensation, superfluidity}
\label{sec:lasing-cond-superfl}

Condensation, coherence and superfluidity are often connected,
however when dealing with two-dimensional systems of composite
particles, where finite-size and non-equilibrium effects may be
important, it is important to separate and clarify these concepts.  
The discussion below compares condensation, coherence and
superfluidity in equilibrium infinite systems to the case including
finite-size and non-equilibrium effects.  
For simplicity we discuss these ideas in terms of a general bosonic
field $\psi^{\dagger}$, rather than any specific field appearing in
the microcavity polariton system.
The first concept is macroscopic occupation of single-particle
wave-function; rigorously this can be defined as the existence of a
macroscopic eigenvalue of the reduced one-particle density
matrix~\cite{penrose56}:
\begin{equation}
  \label{eq:one-pcl-rho}
  \rho_1(\vect{r},\vect{r}^{\prime}) = \left< \psi^{\dagger}(\vect{r})
  \psi^{}(\vect{r}^{\prime}) \right> = \sum_i n_i
  \varphi_i^{\ast}(\vect{r}) \varphi_i^{}(\vect{r}^{\prime}),
\end{equation}
where $n_i$ is the occupation of the single-particle mode
$\varphi_i(\vec{r})$.
A macroscopic eigenvalue exists if $\lim_{N\to\infty} n_0/N \ne 0$ where
$N = \sum_i n_i$ is the total number of particles.
In an infinite system, if the macroscopically occupied state is an
extended state, then there is \emph{Off Diagonal Long Range Order} ---
i.e.\ if in the position representation $\lim_{\vect{r} \to \infty}
|\varphi_0(\vect{r})| \ne 0$, then there remain extensive terms far
from the diagonal~\cite{yang62}.
Thus, in such a case, the correlation function $\lim_{\vect{r} \to
\infty} \left< \psi^{\dagger}(\vect{r}) \psi{}(0) \right> = n_0
\varphi_0^{\ast}(\vect{r}) \varphi_0^{}(0) \ne 0$.
In a non-interacting two-dimensional trapped (and thus finite) gas of
bosons there can be a sharp crossover\footnote{In the limit of
  vanishing trap curvature and infinite number of particles the
  crossover becomes a phase transition, but only if trap curvature
  vanishes as the correct power of the number of
  particles~\cite{bagnato91,ketterle96}.}  to a state with a
macroscopic eigenvalue (i.e.\ of the order of the number of particles)
of the one-particle density matrix~\cite{bagnato91,ketterle96}.
However, the single-particle state $\varphi_0(\vect{r})$ to which this
eigenvalue corresponds will be a state localised in the trap, and so
despite the existence of a macroscopic eigenvalue, $\lim_{\vect{r} \to
\infty} \left< \psi^{\dagger}(\vect{r}) \psi{}(0) \right> = 0$.

The visibility of interference fringes is directly related to
the first-order coherence function $g_1(t=0;\vec{r})$, where:
\begin{equation}
  \label{eq:define-gtr}
    g_{1}(t;\vec{r}) = \frac{%
    \left< \psi^{\dagger}(\vect{r},t) \psi^{}(\vect{0},0) \right>
  }{%
    \sqrt{%
      \left< \psi^{\dagger}(\vect{0},0) \psi^{}(\vect{0},0) \right>
      \left< \psi^{\dagger}(\vect{r},t) \psi^{}(\vect{r},t) \right>
    }
  };
\end{equation}
coherence can be defined by the properties of this function.
As just discussed, if one defines coherence by the limit of
$g_1(0,\vect{r}\to\infty)$, then in a trapped system, this function
vanishes. 
However, coherence will exist across the size of the trap.
In such inhomogeneous and complicated cases, a binary
classification of coherent/incoherent is less useful than a
description of how the coherence varies as a function of separation in
time and space.
Both the size of this variation, and its functional form will depend
on the interplay of finite size (and form of trapping potential),
temperature, interactions, and decay rates~\cite{szymanska06:long}.

Superfluidity can meanwhile be defined separately as the difference of
longitudinal and transverse response functions at vanishing
wave-vector~\cite{khalatnikov,popov,pitaevskii03}:
\begin{eqnarray}
  \nonumber
  \chi_{ij}(\omega=0,\vect{q})
  &=&
  2 \int_0^{\beta} d \tau
  \left<\left<
      J_i(\vec{q},\tau) J_j(-\vec{q},0)
    \right>\right> \\
  &=&
  \chi_T(q) \left( \delta_{ij} - \frac{q_i q_j}{q^2} \right)
  +
  \chi_L(q) \frac{q_i q_j}{q^2}\; ,
\label{eq:superfluidity}
\end{eqnarray}
and $\rho_s \propto \lim_{q\to 0} \left[\chi_L(q) - \chi_T(q)
\right]$.
Since superfluidity results from a change of the response functions,
it occurs only in an interacting system; without interaction
  bosons do not become superfluid.
In a two-dimensional infinite interacting system, below the BKT
transition~\cite{nelson77,kosterlitz73,fisher88}, coherence decays as
a power law rather than an exponential --- low energy phase
fluctuations prevent true long range order~\cite{mermin66}
\footnote{Above the BKT transition, in addition unbound vortices are
  present, and these lead to exponential decay of correlations.} ---
and superfluidity exists, but no macroscopic occupation of a single
mode.

\begin{figure}[htpb]
  \centering
  \includegraphics[width=0.8\linewidth]{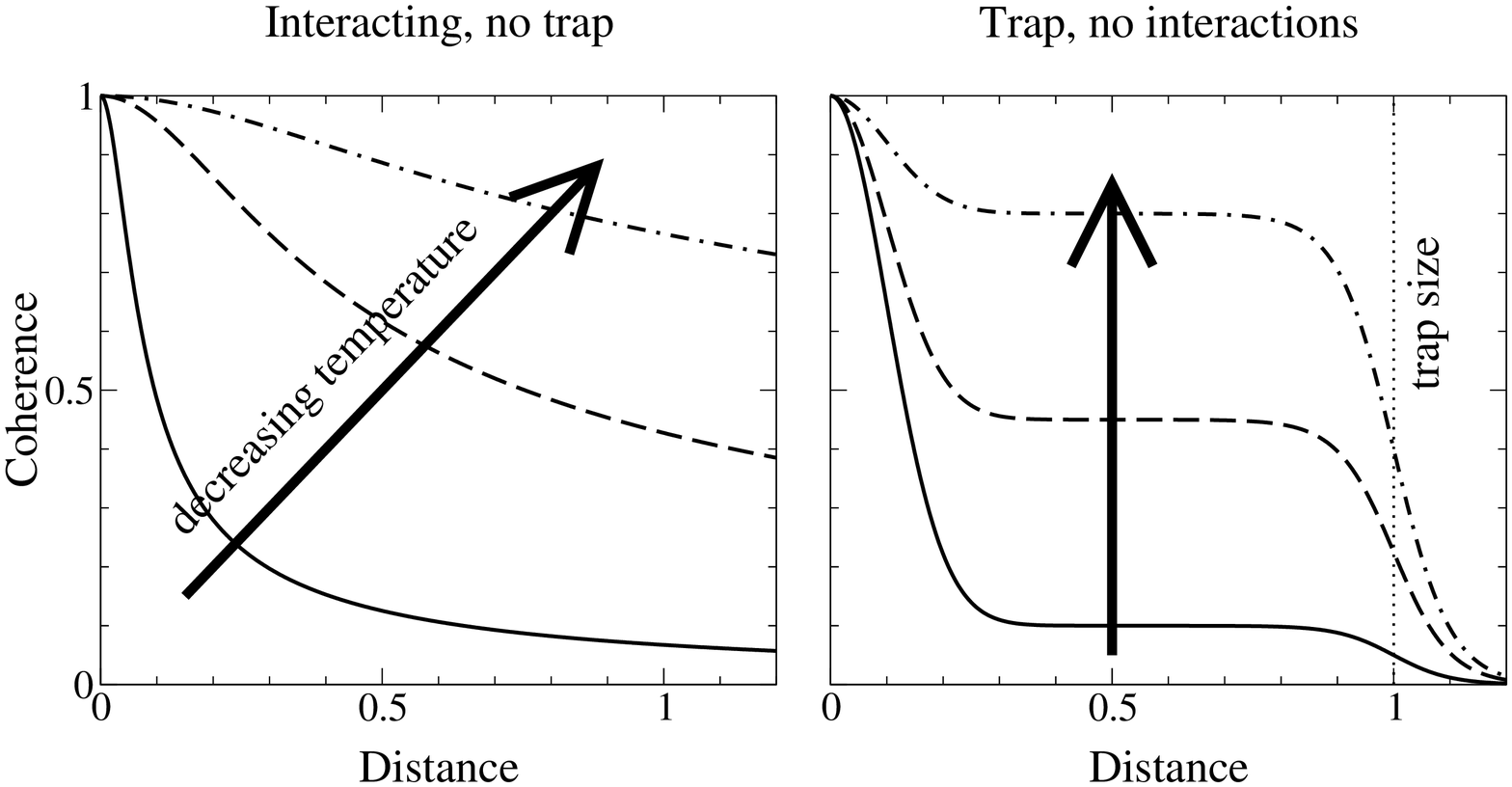}
  \caption{Sketches of the spatial decay of coherence,
    $g_1(t=0;\vec{r})$.  Left: interacting infinite 2D Bose gas,
    showing power-law decay at long distances.  Right: non-interacting
    trapped gas; at short distances (less than the thermal length)
    perfect coherence exists; on intermediate lengthscales coherence
    reaches a constant value, due to the non-zero condensate fraction;
    coherence eventually goes to zero for separations larger than the
    trap size.}
  \label{fig:coherence}
\end{figure}

When one considers a more realistic system, which is both interacting,
but also of finite extent, one cannot ignore a-priori the physics of
the BKT transition,
nor can one ignore
a-priori the physics of the trap.
At low enough temperatures it is clear there will be macroscopic
occupation of a single mode, and full coherence across the trap.
How this state is approached as temperature is reduced, or
as density is increased differs depending on whether
interactions or finite size effects are dominant.
If described as a non-interacting gas, the coherence at all distances
increases uniformly as a single mode is increasingly
occupied~\cite{bagnato91,ketterle96}. 
In the BKT scenario, power law correlations develop on intermediate
scales (between some short range thermal length and the trap size);
then as temperature decreases, the thermal length increases and the
power with which correlations decay decreases, again restoring full
coherence as $T\to 0$~\cite{petrov00,keeling04:angular},
as shown in Fig.~\ref{fig:coherence}.

\begin{figure}[htpb]
  \centering
  \includegraphics[width=0.6\linewidth]{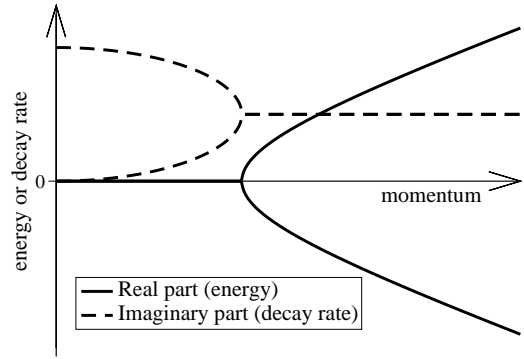}
  \caption{ Sketch of momentum dependence of real and imaginary parts
    of poles of the Green's function, in the presence of pumping and
    decay.
    At zero momentum, there is a free phase, i.e.\ a mode with
    vanishing real and imaginary part.
    For small momentum both modes
    are diffusive, with no real part, but a non-zero imaginary part.
    Only above some critical momentum does a real part develop. 
    Such results are seen both in the calculations in
    Refs.~\cite{szymanska06:prl,szymanska06:long}, and also for
    resonant pumping (as discussed in
    Sec.~\ref{sec:phase-degree-freedom}) in
    Refs.~\cite{wouters05,wouters06}.}
  \label{fig:diffusive-poles}
\end{figure}

Adding non-equilibrium effects, the nature of decay of correlations in
an infinite system is significantly
altered~\cite{szymanska06:prl,szymanska06:long}.
The long wavelength phase modes, responsible for decay of correlations
become diffusive.
i.e.\ the poles of the Green's functions, which in the equilibrium case
have the form $\omega \simeq \pm c k$, take instead the form $\omega
\simeq i x \pm \sqrt{(ck)^2 - x^2}$ in the pumped and decaying case.
This is illustrated in Fig.~\ref{fig:diffusive-poles}.
Combining the effects of phase diffusion, and discrete level
spacing~\cite{szymanska06:long} the properties of coherence are further
modified, and one approaches the laser limit: temporal coherence of
laser emission comes from one, or at most a few modes (resonant with
the laser cavity), and so phase diffusion of a single
mode~\cite{haken70} leads always to exponential decay of temporal
correlations~\cite{holland96,tassone00,porras03}.
As well as this distinction of forms of temporal coherence, it is
worth mentioning here a few other important distinctions between
lasing and the generalised concept of condensation discussed here, as
there are evidently also similarities~\cite{haken75}.
Most obviously, polariton condensation is seen at the polariton
resonance, which is significantly (of the order of the Rabi splitting
$\Omega_R$) below the lowest cavity photon mode; as such nonlinear
emission coexisting with strong coupling is a signal that one should
consider polaritons, and not just photon lasing.
A more fundamental difference is that lasing requires inversion, while
condensation does not; this is a consequence of the standard laser
systems possessing little coherence in the gain medium, while
excitons, being part of a coherent polaritons, are
coherent~\cite{szymanska02,szymanska03:pra,eastham03:ssc,marchetti04:prb,marchetti04:comparison}.
This distinction can also be seen by comparing the critical lasing
condition to the Gross-Pitaevskii equations of Eq.~(\ref{eq:gpe}) and
Eq.~(\ref{eq:gpe-fermi}).
In the presence of pumping and decay, the susceptibility
  (describing the nonlinear response of excitons) becomes
  complex~\cite{szymanska06:prl,szymanska06:long}; the real part of the
  susceptibility gives the nonlinearity in the Gross-Pitaevskii
  equation.
In contrast, the imaginary part of the susceptibility describes
absorption or gain, and leads to the lasing condition, that round trip
gain and loss balance.
A treatment of a model system with pumping and decay elegantly shows
how these conditions can be combined, giving an expression in terms of
the total susceptibility~\cite{szymanska06:prl,szymanska06:long}.
 
Starting from the strong-coupling regime, when crossing over to
the weak-coupling regime (see, e.g.~\cite{houdre95,cao97,butte02}),
the polariton splitting collapses, and so the lasing mode no longer
has any excitonic character, and becomes the standard photon laser.
A major advantage of an exciton-polariton laser over standard lasers
is that it can operate without the inversion of the electronic
population~\cite{imamoglu96}, and therefore it has a much
smaller threshold pump power.
It is interesting to note that wide band gap semiconductors, such as
GaN and ZnO, would be particularly suitable as in these cases excitons
are stable at higher temperatures and densities, and therefore they
could operate at room temperature.
Electronic population inversion is not necessary because in the
exciton-polariton laser, both photon field and excitons are
coherent.
In addition, the involvement of the excitonic field leads to strong
nonlinear effects compared to conventional lasers, due to
exciton-exciton interactions.

\subsection{Phenomena}
\label{sec:phenomena}

The previous two sections discussed the models, and the treatments
of the environment, that have been used to theoretically model
polariton condensates.
This section in contrast will review a few of the phenomena that
have been predicted as possible signatures and properties of
a polariton condensate.

\subsubsection{$T_c$ and the phase boundary}
\label{sec:t_c-phase-boundary}

Within a given model, and effective description of the environment,
it is natural to first ask under what conditions a condensate
can exist.
Within an equilibrium model of the lower polariton branch as weakly
interacting bosons, phase diagrams for the physical parameters of
various possible materials are shown in
Refs.~\cite{kavokin03:pla,malpuech03} (however, see
Ref.~\cite{keeling06} for a discussion of the effects of non-quadratic
dispersion on the BKT transition temperature).
By considering a simplified version of the Bose-Fermi model
Eq.~(\ref{eq:dicke-model-fermions}), where the energies
$\varepsilon_{\alpha}$ are described by a Gaussian distribution, while
all excitons display a fixed coupling to light, the mean-field phase
boundary was first calculated in Refs.~\cite{eastham00:ssc,eastham01}.
The effect of fluctuations, restoring the bosonic limit at low
densities was instead considered in
Refs.~\cite{keeling04:polariton,keeling05}.
 Since the content of the boson-fermion model at small densities and
 temperatures is equivalent to a bosonic model, and the low momentum
 part of the polariton dispersion is controlled by the photon mass, it
 is not surprising that it is possible to recover the standard
 BKT transition temperature of a weakly
 interacting Bose gas from the boson-fermion model in the low density
 limit.
A calculation of the mean-field boundary
which instead takes into account a realistic
description of the quantum well disorder and the full distribution of
oscillator strengths (see Fig.~\ref{fig:allg}) has been performed in
Refs.~\cite{marchetti06:prl,marchetti06:rrs_long} (see
Fig.~\ref{fig:phased}).

Owing to finite size effects the experimental systems do not
have a sharp phase transition marking the onset of a broken symmetry.
All the observed transitions are rounded, and in order to extract a
phase boundary from experiment, some criterion has to be chosen. 
One commonly used criterion is the nonlinear threshold; i.e.\ the
point at which the relation between emission at $\vec{k}=0$ and input
pump power becomes nonlinear.
Such a criterion is somewhat problematic.
A second-order phase transition can be expected to be accompanied by a
region with large susceptibilities, and thus such nonlinearity extends
over a significant range of parameters, and so identification of a
strict phase boundary from it is hard.
Only in a mean-field theory does the onset of nonlinearity occur
at the transition.
However, because of the long-range nature of interactions, mean-field
theory can be an adequate description for lasers, and for polariton
condensates except at very low densities (as can be seen from
Fig.~\ref{fig:phased}).
The following sections discuss other phenomena that may demonstrate or
describe condensation and coherence in microcavity polariton systems,
and may thus provide alternative, or corroborating experimental
criteria to find the phase boundary.

\subsubsection{Energy-resolved luminescence, resonant Rayleigh scattering}
\label{sec:prop-lumin}

\begin{figure}
  \begin{center}
    \includegraphics[width=0.5\linewidth]{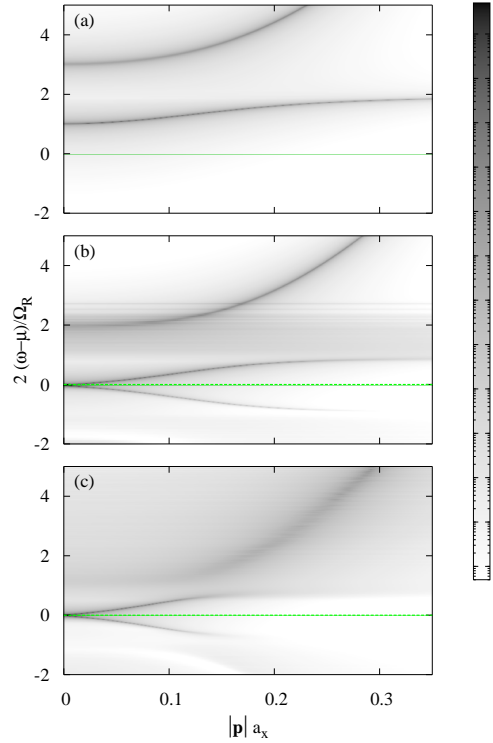}
  \end{center} 
  \caption{ Contourplot of the disorder averaged RRS intensity
    $\left< I_{\vect{p} \vect{q}} (\omega)\right>$ for $|\vect{p}| =
    |\vect{q}|$ as a function of the dimensionless momentum
    $|\vect{p}| a_{\ex}$ and rescaled energy $2(\omega -
    \mu)/\Omega_{\mathrm{R}}$, for zero detuning, Rabi splitting
    $\Omega_{\mathrm{R}}=26$meV, temperature $k_{B} T = 20$K, and a
    disorder strength characterised by an inverse scattering time
    $1/\tau = 1.16$meV. (a) non-condensed (density $\rho\simeq 0$);
    (b) condensed regime 
    ($\rho\simeq 3.6\times 10^{9}\mbox{cm}^{-2}$)
    ; (c) condensed regime 
    ($\rho\simeq 3.1\times 10^{10}\mbox{cm}^{-2}$)
    . Green lines mark chemical potential: if
    non-condensed RRS emission is present only above the chemical
    potential, but when condensed, it is present both above and below.
    [From~\cite{marchetti06:rrs_long}].}
  \label{fig:RRS}
\end{figure}

In both the weakly interacting boson model, and the Bose-Fermi
model, condensation leads to changes in the spectrum of polariton
modes --- most significantly the appearance of the phase
modes~\cite{popov,pitaevskii03}.
In addition, for a model starting from localised exciton states, with
a distribution of energies and oscillator strengths, there is weak
emission from sub-radiant exciton states between the upper and lower
polaritons~\cite{houdre96}.
 This emission is also modified by
  condensation~\cite{marchetti06:prl}: in the presence of a condensate
  these sub-radiant exciton states have energies $E_{\alpha}$ as
  defined following Eq.~(\ref{eq:gpe-fermi}), and so the density of
  states is changed by the coupling to the coherent field.
  This change to the emission is discussed further in detail
  Ref.~\cite{marchetti06:rrs_long}.  
In practice, it is however hard to observe the incoherent luminescence
of thermally excited modes in the presence of a strong signal
from the coherent condensate.
One suggestion to overcome this problem is to probe these excited
modes via resonant Rayleigh scattering; by using a phase sensitive
measurement one may be able to identify a small coherent scattering
signal even in the presence of emission from the
condensate~\cite{marchetti06:prl,marchetti06:rrs_long}.
Fig.~\ref{fig:RRS} shows the Rayleigh scattering signal expected
  both above and below the phase transition; note in the condensed
  case, one sees linear modes both above and below the chemical
  potential; this is as one expects from the Bogoliubov spectrum,
  where the normal modes are superpositions of particle creation and
  annihilation.

\subsubsection{Momentum distribution of radiation}
\label{sec:moment-distr-radi}

By integrating the luminescence at a given wave-vector, one may
consider the momentum distribution of polaritons --- this has also
been discussed in the context of exciton
condensation~\cite{keeling04:angular,zimmermann06}.
In a weakly interacting, two-dimensional, infinite system, the
density of states would be constant, and this would just show the
Bose-Einstein distribution.
In experiments, the distribution at small momentum is close to, but
can deviate from the Bose-Einstein
distribution~\cite{deng03,kasprzak06:nature}.
However, such deviations are expected; both due to interactions, which
modify the spectrum and so modify the density of
states~\cite{keeling04:polariton,keeling05}, and also due to finite
size, which cuts off components at small wave-vectors.
A naive calculation shows how the change of spectrum causes a change
of density of states: for a weakly interacting Bose gas with
dispersion $\varepsilon_\vect{k}$, and mean-field equation $\mu = g\rho_0$, we
have the Bogoliubov spectrum $E_\vect{k} = \pm \sqrt{\varepsilon_\vect{k}(\varepsilon_\vect{k} +
  2 g\rho_0)}$, and simple algebra gives the density of states:
\begin{eqnarray}
  \nonumber \nu_s(\omega,\vect{k}) &=& \Im\left[
  \mathcal{G}^R(\omega+i0^+,\vect{k})\right] = \Im\left[ \frac{i
  \omega + \varepsilon_\vect{k} + g\rho_0}{\omega^2 + E_\vect{k}^2}
  \right] \\ &=& \frac{E_\vect{k} + \varepsilon_\vect{k} +
  g\rho_0}{2E_\vect{k}} \delta(\omega-E_\vect{k}) + \frac{E_\vect{k} -
  \varepsilon_\vect{k} - g\rho_0}{2E_\vect{k}}
  \delta(\omega+E_\vect{k})\; .
\label{eq:naive-dos}
\end{eqnarray}
This suggests that as $\vect{k} \to 0$, the density of states goes like $g
\rho_0/E_\vect{k} \propto 1/k$, however this neglects the fact that the low
energy modes are phase modes, and phase fluctuations may grow without
bounds, as only their gradient costs energy; a full calculation,
e.g.~\cite{marchetti04:prb,keeling05}, shows that this
term becomes $1/k^{2-\eta}$ with $\eta \simeq 2 T/T_{BKT}$.
In order to include the effects of finite size, one approach is to
start with the zero-temperature Thomas-Fermi spatial profile, for
which coherence across the whole cloud leads to sharp angular peaks,
and then consider how phase fluctuations destroy the long range
coherence, and thus soften the peaks~\cite{keeling04:angular}.
There have also been calculations of this momentum distribution for
non-equilibrium situations, where details of the dynamics of polariton
relaxation lead to a maximum of the distribution at non-zero $\vect{k}$, e.g.\
Refs.~\cite{tassone97:prb,porras02,doan05:prb,cao04}.

\subsubsection{Linewidth, first- and second-order coherence, polarisation}
\label{sec:lumin-linew}

The line-shape of emission is controlled by the Fourier transform of
the first-order coherence function as a function of time, i.e.
$g_1(t;\vec{r}=0)$, using $g_1(t;\vec{r})$ defined in
Eq.~(\ref{eq:define-gtr}).
A perfectly coherent single mode source would have $g_1(t;0)$ constant,
and thus infinitely sharp lines.
Although at any finite temperature, population of slow phase modes
will lead to some decay of $g_1(t;0)$, the transition to a condensed
phase will lead to a slower decay of $g_1(t;0)$\footnote{One should note
in defining a coherence time that, for an equilibrium, infinite,
two-dimensional system, the long time decay would be power law, and so
there is no well defined coherence time, however finite size effects
lead to exponential decay~\cite{szymanska06:long}.}, as has been
seen~\cite{kasprzak06:thesis}.
Similarly, as discussed in Sec.~\ref{sec:lasing-cond-superfl}, one can
also consider the spatial decay of coherence~\cite{kasprzak06:nature};
which has also been much studied in cold atomic gases (see,
  theoretical discussion in
  Refs.~\cite{glauberPRA99,petrov00,demlerPNAS06} and experiments in
  Refs.~\cite{hagley99,bloch99,hadzibabic06}).

The calculation of coherence has already been discussed briefly in
Sec.~\ref{sec:lasing-cond-superfl}.
As was stressed there, a distinction exists between coherence of a few
mode laser~\cite{holland96}, applied to the polariton problem in
Refs.~\cite{tassone00,porras03}; and coherence in a continuum of
modes~\cite{mieck02,szymanska06:prl,szymanska06:long}.
In both cases increasing temporal coherence in the system should
lead to a narrowing of linewidth as the phase transition is
approached, however the behaviour far above the transition may differ.
This is clear from the fact that for a single mode with a quartic
interaction, the equilibrium state is a number state, and so starting
from a coherent state, one has phase diffusion due to \emph{self phase
modulation} even in the absence of any pumping or
decay~\cite{porras03}.
For a many-mode system, the ground state is neither a coherent nor a
number state, and is better described by the Bogoliubov-Nozi\`eres
state~\cite{nozieres82}, thus it is not clear that the same effect
--- i.e.\ \emph{self phase modulation} causing larger broadening ---
should persist.

A related measurement is the second-order coherence
function (see e.g.\ Ref.~\cite{gardiner}):
\begin{equation}
  \label{eq:second-order-coherence}
  g_2(t) = \frac{%
    \left< \psi^{\dagger}(\vect{r},0) \psi^{\dagger}(\vect{r},t) \psi^{}(\vect{r},t) \psi^{}(\vect{r},0) \right>
  }{%
    \left< \psi^{\dagger}(\vect{r},0) \psi^{}(\vect{r},0) \right>
    \left< \psi^{\dagger}(\vect{r},t) \psi^{}(\vect{r},t) \right>
  }
\end{equation}
As mentioned above, for a thermal state, $g_2(t=0)=2$, while for a
coherent state $g_2(t=0)=1$.
Experimental measurement of $g_2(t=0)$ is restricted by finite
detector integration times, and since $g_2(t)=1$ at times when
$g_1(t;0)\to 0$, it is hard to distinguish the value of
$g_2(t=0)$~\cite{deng02,kasprzak06:thesis}.
The dynamical behaviour of $g_2(t=0)$ as a function of time
following switching on the pump laser~\cite{rubo03,rubo04,laussy04:prl},
or from an equilibrium picture with separate coherent and incoherent
contributions~\cite{laussy06} has been considered.
Within the Bose-Fermi model, the second-order coherence has been
studied for a finite number of excitons coupled to a single
coherent field, where the finite number of states replaces the phase
transition with a smooth crossover~\cite{eastham06}.

Considering the differences of interaction strength between polaritons
with parallel and anti-parallel spin polarisations, Laussy {\it et
  al.}~\cite{laussy06} have shown that condensation should be associated
with spontaneous development of a linear polarisation.
That a condensate should choose some definite polarisation state is a
much more general result, i.e.\ that repulsive interactions prevent
fragmentation of a condensate, even when there are two single-particle
states with identical energies~\cite{nozieres}.
Considering the specific form of the interaction, Laussy {\it et al.}
showed that this specific polarisation should be a linear state (as
opposed to circular or elliptical).
Note, however, that the pinning of the polarisation to one of the
crystallographic axes observed in
Refs.~\cite{kasprzak06:nature,klopotowski2006,amo2005} cannot be an
effect of spontaneous symmetry breaking, but it is instead likely due
to some optical anisotropy of the microcavities.
In atomic gases, investigation of ``spinor condensates'' is hard, as
it requires the use of all optical trapping, as a magnetic field would
Zeeman split the atomic spin states; further the populations of spin
species are effectively fixed at the start of the experiment, so
rather than polarisation, phase separation into spin domains has been
observed there~\cite{stenger98}.

\section{Resonant pumping}
\label{sec:resonant-pumping}
Since its first realisation in 2000~\cite{savvidis00:prl}, the
possibility of reaching the stimulated scattering regime for
polaritons by resonant pumping has attracted considerable
interest~\cite{savvidis00:prb,baumberg00:prb,houdre00,stevenson00,huang00,messin01,saba01,huynh03,kundermann03,butte03,langbein04,diederichs06,baas06:prl},
and has initiated the search for polariton lasers and a new generation
of ultrafast optical amplifiers and switches.
In resonant excitation experiments, polaritons are optically pumped at
an energy and momentum which allows coherent polariton-polariton
scattering directly into the ground state.
In pulsed, ultrafast pump-probe experiments, a second pump is used to
initiate the stimulated scattering process to the ground state, while
for c.w.\ (continuous wave) excitation the stimulated regime can be
self-induced above some pump threshold.
As in non-resonant excitation experiments, the direct mapping between
in-plane momentum and angle of bulk photons is crucial.
It is this direct mapping that allows one to directly pump at a given
momentum, and thus to perform the experiments described here.

The macroscopic occupation of the ground state in these experiments
differs in some ways from condensation and spontaneous coherence arising
from incoherent pumping.
In experiments without a probe pulse, there is a free phase between
the pump and signal modes, however as discussed in
Sec.~\ref{sec:phase-degree-freedom}, the existence of a free phase
alone is not sufficient to ensure superfluidity in such driven
systems.
Further, since the coherence of the signal beam is directly
coupled to the pump beam, higher order correlations, and the
linewidth of the signal may be inherited from the pump.
Experiments with an additional probe beam at the signal frequency
differ yet further; in such experiments the phase of the signal is
fixed by the probe, and so in this case all
excitations ought to be gapped.

It is interesting to note that the idea of parametric scattering
discussed here for polaritons has been recently applied in a
different field; that of dilute gases of atoms confined in an optical
lattice~\cite{deng99,campbell06}.
We will discuss this example in more detail later in
Sec.~\ref{sec:feshb-reson-ferm}.

\begin{figure}
\begin{center}
\includegraphics[width=1\linewidth]{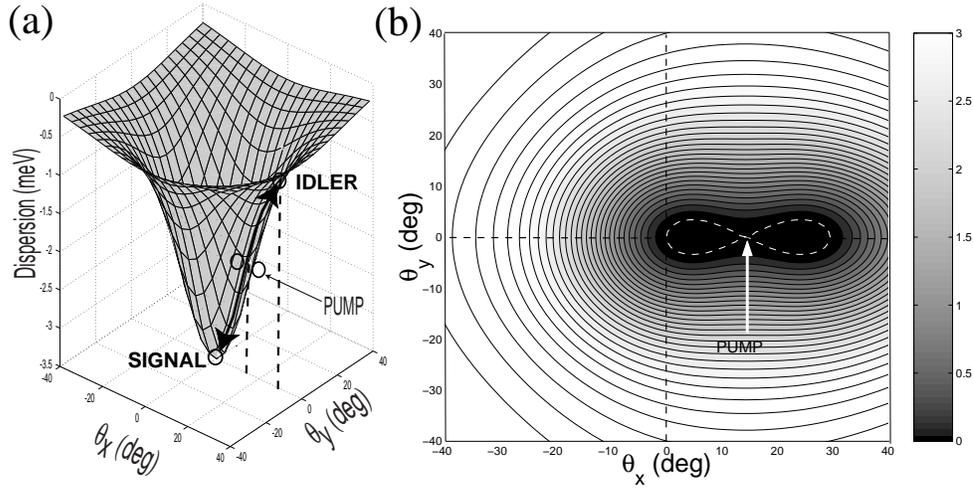}
\end{center} 
\caption{\small (a) Lower polariton dispersion; (b) countorplot of
  $|E^{\rm LP}_{\vect{k}} + E^{\rm LP}_{2\vect{k}_p-\vect{k}} - 2
  E^{\rm LP}_{\vect{k}_p}|$ as a function of $\vect{k}$ and the zero
  value contour (white dashed). A pair of final states (signal and
  idler) can be found by intersecting the white dashed curve with a
  straight line passing through $\vect{k}_p$ (indicated as pump).
  [From~\cite{ciuti01} Copyright (2001) by the American Physical Society].}
\label{fig:magicangle}
\end{figure}
%
\subsection{Summary of experiments}
\label{sec:summary-experiments-res}
The main idea of resonant pumping
experiments~\cite{savvidis00:prl,savvidis00:prb,baumberg00:prb,houdre00,stevenson00,huang00,messin01,saba01,huynh03,kundermann03,butte03,langbein04,diederichs06,baas06:prl}
is that of the coherent scattering of two polaritons from the
resonantly pumped mode (pump) into the ground state (signal) and a
high energy state (idler) [see Fig.~\ref{fig:magicangle}(a)].
Energy and momentum conservation in this scattering requires one
to have  $\{\vect{k}_p, \vect{k}_p\} \mapsto \{0,2\vect{k}_p\}$, where
\begin{equation}
  E^{\rm LP}_{\vect{k}=0} + E^{\rm LP}_{2\vect{k}_p} = 2 E^{\rm
  LP}_{\vect{k}_p} \; ,
\end{equation}
which uniquely selects the momentum (i.e.\ angle) of pump, signal and
idler.
If one instead relaxes the condition $\vect{k}=0$ for the signal mode,
then for a fixed pump angle $\vect{k}_p$, it can be
shown~\cite{ciuti01,langbein04} that those final states which satisfy
energy and momentum conservation describe a figure-of-eight in
momentum space [see Fig.~\ref{fig:magicangle}(b)].
The non-parabolic dispersion of polaritons is crucial in order to
obtain such a resonant scattering processes, and so the analogous
process for excitons is forbidden.
Resonant experiments can be divided into two types: In the first
type the scattering is stimulated by a weak probe field (parametric
amplification), while in the second type there is no probe and the
seed to initiate stimulation is provided by the pump itself, if above
a certain power threshold (parametric oscillation).

\begin{figure}
\begin{center}
\includegraphics[width=0.7\linewidth]{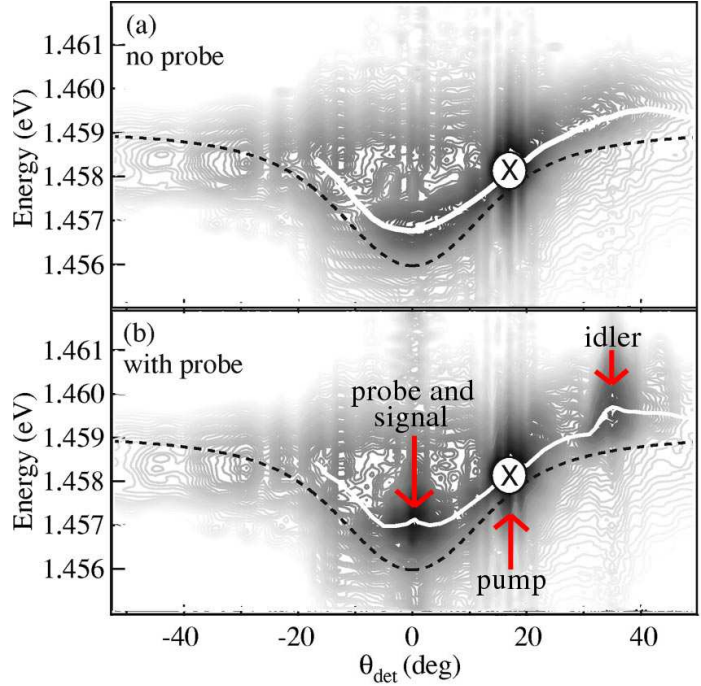}
\end{center} 
\caption{\small  Contour maps of
  the PL emission spectra for excitation at the `magic' angle
  ($\times$) (a) when no probe is applied at zero angle and (b) when a
  probe is applied, which stimulates a strong gain in the probe
  direction. The peak of the PL (white line) shows a marked asymmetry
  and a blue-shift of $\sim 0.7$meV with respect to the low power
  polariton dispersion (dashed). [From~\cite{savvidis00:prb} Copyright
  (2000) by the American Physical Society].}
\label{fig:probenoprobe}
\end{figure}

In their pioneering work~\cite{savvidis00:prl}, Savvidis and
collaborators realised the stimulated scattering regime for the first
time in an InGaAs/GaAs/AlGaAs microcavity with a Rabi splitting of
$\Omega_R = 7$meV.
By pumping polaritons at the ``magic'' angle $\theta \simeq
17^{\circ}$ (for zero exciton--photon detuning) close to the
inflection point of the LP dispersion curve, and using a second pump
at zero angle (probe) to initiate the process, a substantial gain of
up to 70 was observed.
The large observed signal required both the ``magic'' angle
pumping, and the probe at zero angle, and was absent either if no
probe was applied~\cite{savvidis00:prb} (see
Fig.~\ref{fig:probenoprobe}), or when pumping at different angles even
in presence of the zero angle probe.
These results provide strong evidence for a polariton scattering
process stimulated by the probe.
The bosonic scattering rate is enhanced by a factor $N+1$ where $N$ is
the population of bosons in the final state~\cite{scully97}, in this
case the ground state.
Moreover, stimulated scattering was confirmed in this experiment by
the observed exponential dependence of the gain on pump power.
Note however that the obtained stimulated regime is more correctly
described as a `parametric' amplifier rather than a laser, i.e.\ the
population of the lowest mode is amplified by a phase coherent
parametric process~\cite{messin01}.

Much experimental work has followed this first
result~\cite{savvidis00:prb,baumberg00:prb,houdre00,stevenson00,huang00,messin01,saba01,huynh03,kundermann03,butte03,langbein04,diederichs06,baas06:prl}.
Evidence of stimulated scattering has been obtained in~\cite{huang00},
with a three beam pulsed experiment, where two pumps excite states at
large and opposite angles, $\theta_p \sim \pm 45^{\circ}$, and a third
beam is used as a probe at normal incidence.
Here, the pump creates a quasi-thermal exciton reservoir at large
momentum; then, scattering of polariton from $\vect{k}_p$ and
$-\vect{k}_p$ to the $\vect{k}=0$ upper and lower polariton states can
occur, and is is stimulated by the occupation (due to the probe) of
the final state.
Note that, at zero detuning, such a scattering process conserves
energy ($2\varepsilon^{\rm ex}_{\vect{k}} = E^{\rm LP}_0 + E^{\rm
  UP}_0$) and momentum.
Recent experiments in the original two beam pump probe configuration,
but with time resolved measurements as a function of different pump
and probe angles have begun the investigation of the hydrodynamic
properties of the injected ``polariton
fluid''~\cite{baumberg05,lagoudakis06}.

In another series of experiments, making use of resonant c.w.\
excitation~\cite{stevenson00,baumberg00:prb} rather than ultrafast
pulsed excitation as in~\cite{savvidis00:prl,savvidis00:prb,huang00},
the stimulated scattering regime has been reached even without
the probe beam (parametric oscillation).
Here, the stimulated scattering is self-initiated when the pump power
is strong enough that the final state population becomes close to one.
It is interesting to note that, as in the non-resonant excitation
experiments, a minimum pump power is required (with threshold-like
behaviour) in order to overcome the `bottleneck' effect and thus
provide the occupation of the ground state required to stimulate
scattering.
In these particular experiments, occupation of the ground state was
estimated to be close to $1$ at threshold and around $300$ at the
highest pump powers.
A maximum blue-shift of the lower polariton of the order of $0.5$meV
was observed, which is much less than the Rabi splitting of $\sim
6$meV, and so confirms that the experiment is always in the strong
coupling regime.
Because of the stimulated scattering, pumping of the ground state mode
is efficient, and so these systems have a low power threshold,
typically $5$ times smaller than for a high quality vertical cavity
surfaces emitting laser (VCSEL)~\cite{baumberg00:prb}.

While the experiments described above were at temperatures of the
order of $5$K, Saba and collaborators have reported pump-probe
parametric amplification of polaritons with an extraordinary gain up
to $5000$ at temperatures up to $120$K in GaAlAs-based microcavities,
and up to $220$K in CdTe-based microcavities~\cite{saba01}.
The highest possible operating temperatures for observing amplification
have been shown to be determined by the exciton binding energy
($25$meV for the used CdTe wells and $13.5$meV for GaAs wells), rather
than the polariton Rabi splitting, which in these experiments varies
from $25$meV for the 24 quantum-well (QW) CdTe microcavity to
$15$meV and $20$meV for GaAlAs microcavities with respectively 12 QWs
and 36QWs.
Time-resolved measurements, obtained by controlling the delay between
pump and probe, show ultrafast dynamics of the parametric gain,
promising future applications in high-repetition-rate optical switches
and amplifiers.

As an alternative to the finite angle resonant configuration, a few
experiments~\cite{messin01,diederichs06} have concentrated on the
double energy and momentum resonance, where pump, signal and idler are
all at normal incidence, $\vect{k}_p=0$.
In this geometry it is possible to investigate the dependence of the
phase of the signal on the phase of the pump, and thus to
show~\cite{messin01} that in this degenerate configuration the
parametric scattering is a coherent process.
In addition, in most of the early resonant experiments, such as
Refs.~\cite{savvidis00:prb,baumberg00:prb,stevenson00,saba01}, 
spectral narrowing of the signal was observed when amplification
occurs, also suggesting that the signal is
coherent~\cite{kundermann03}.
Direct evidence for the coherent nature of the signal emission has
come only recently, in Ref.~\cite{baas06:prl}, where first-order
temporal and spatial coherence were investigated.
In that work, in a resonant c.w.\ pump, no probe, configuration, two spots
separated by $70\mu$m, coming from the same laser excited region of
the sample ($100\mu$m FWHM of Gaussian profile), are overlapped in
momentum space, showing interference fringes.
In addition, by making use of noise measurements, the emitted signal
is shown to be in a single-mode quantum state, rather than in a
multi-mode state.

Recently, the pair correlation of the emitted signal-idler polaritons
have been demonstrated by showing that polaritons in two distinct
idler modes can interfere if and only if they share the same signal
mode~\cite{savasta05:prl}.

\subsection{Theories}
\label{sec:theories}

Predictions about threshold conditions, spectral properties and
efficiency of the amplification in resonant pump-probe experiments can
be easily obtained by making use of the effective lower polariton
Hamiltonian $H_{\rm LP}$~(\ref{eq:bosonic-lp}) described in
Sec.~\ref{sec:weakly-inter-boson}, to which one must add the
coupling to the external radiation pump $\Omega_{\rm pump} (t)$ and
probe $\Omega_{\rm probe} (t)$ fields~\cite{ciuti00:prb}:
\begin{equation}
  \label{eq:external-driving-hamiltonian}
  H_{\rm ext} = c \sum_{\vect{k}}
  \left\{\left[\delta_{\vect{k},\vect{k}_p} \Omega_{\rm pump} (t) +
  \delta_{\vect{k},0} \Omega_{\rm probe} (t)\right] \cos
  \theta_{\vect{k}} L_{\vect{k}}^\dag + \rm{H.c.}\right\} \; .
\end{equation}
A closed set of equations of motion for the expectation values of
probe (signal) $\left< L_{0}(t) \right>$, pump $\left< L_{\vect{k}_p}(t)
\right>$, and idler $\left< L_{2\vect{k}_p}(t) \right>$ modes can be
obtained by factorising field expectation values, and neglecting higher
order correlations.
By solving such equations, both in the steady-state regime and
numerically for the pulsed excitation, Ciuti and
collaborators~\cite{ciuti00:prb} have obtained the conditions for gain
threshold, showing that the efficiency of the amplifier depends very
strongly on the polariton linewidth; the larger the linewidth, the
higher the threshold and the lower the maximum gain.
Similar results, such as the blue-shift of the signal with increasing
pump power, have been obtained by Whittaker~\cite{whittaker01:prb}
in a classical nonlinear optics treatment.
In Whittaker's paper, a phenomenological model --- where a nonlinear
excitonic oscillator is coupled to the cavity mode, driven by external
fields --- can describe both parametric amplification and oscillation.
This treatment, considering classical fields, and retaining only the
frequencies corresponding to pump, signal and idler modes, is
equivalent to factorising the field expectation values and considering
the equation of motion for $\left< L_{0}(t) \right>$, $\left<
L_{\vect{k}_p}(t) \right>$, and $\left< L_{2\vect{k}_p}(t) \right>$.

By considering the case of a c.w.\ pump without a probe, and expanding
up to the second-order in the field expectation values, it can be also
shown~\cite{ciuti01,ciuti03} that, below threshold for parametric
amplification, the polariton photoluminescence,
\begin{equation}
  \label{eq:photoluminescence-definition}
  {\rm PL} (\vect{k},t,\omega) \simeq \cos^2 \theta_{\vect{k}} \Re
  \int_0^\infty d\tau e^{-i (\omega - i0^+) \tau} \left<
  L_{\vect{k}}^\dag (t+\tau) L_{\vect{k}} (t)\right> \; ,
\end{equation}
has a blue-shifted asymmetric emission distribution, as was observed
in~\cite{savvidis00:prb}, and shown in Fig.~\ref{fig:probenoprobe}.
The lower polariton blue-shift here, $E^{\rm LP}_{\vect{k}} \mapsto
E^{\rm LP}_{\vect{k}} + V^{\rm eff}_{\vect{k},\vect{k}_p,0} \left<
L_{\vect{k}_p}\right>^2$, is due to the interaction term of
Eq.~(\ref{eq:effpot}).
Note that the above formalism is valid only below the threshold for
parametric emission, as above threshold the equation of motion for the pump
mode, describing pump depletion should also be included.

Recent theoretical work~\cite{whittaker05}, including the effects of
polariton blue-shifts on the parametric oscillator equations, has
shown that in the c.w.\ configuration the `magic' angle is not
necessary, as it is in the ultrafast pump-probe case, and that, under
suitable pump conditions, the parametric oscillator can in general be
observed for pump angles $\theta \gtrsim 10^{\circ}$.
Such calculations seem to explain the experimental results obtained in
Ref.~\cite{butte03}.
In that experiment, with c.w.\ pumping, stimulation was achieved over
a wide range of pump angles, from $10^{\circ}$ to $24^{\circ}$ and
with the signal always at $\vect{k}=0$.
Including the polariton blue-shift also allows one to distinguish
parametric oscillation, where output intensity grows continuously, and
bistability, where output suddenly jumps at some threshold pump power,
and shows hysteresis if power is then reduced.
The regimes of bistability and parametric oscillation have been
investigated as a function of pump angle for resonant
pumping~\cite{whittaker05}, and also allowing for mismatch between
pumping frequency and the polariton energy at the pumping
angle~\cite{wouters06b}.

There has also been a proposal to study the hydrodynamic properties of
the injected polariton fluid by studying its coherent scattering by
disorder~\cite{carusotto04}; i.e.\ resonant Rayleigh scattering.
In the proposed experiment, a strong pump beam creates a large
coherent population of polaritons, and also provides
the source of polaritons which may be coherently scattered:
In the presence of disorder, the polaritons can resonantly scatter to
states with different momentum, but the same energy.
This can be observed by looking at the angular distribution of photons
escaping the cavity which are resonant with the pump beam.
At low pump power, this emission pattern will be a ring, at the pump
angle --- i.e.\ those states with the same energy have the same modulus
of momentum.
However, since a large coherent population of polaritons modifies the
polariton dispersion relation, at higher pump powers, both the shape
and intensity variation across the pattern of resonantly scattered
photons reveal information about the polaritons in the cavity.

\subsubsection{Phase degree of freedom and low energy modes}
\label{sec:phase-degree-freedom}

The laser and an equilibrium polariton condensate form extreme ends of
the spectrum of systems in which coherent emission results from a
symmetry breaking transition; the resonantly pumped polariton laser
falls somewhere in between.
Although, as discussed in Sec.~\ref{sec:summary-experiments-res}, the
coherence of the signal is inherited from the pump, in the parametric
oscillator configuration, without a seed signal beam there is in
principle a free phase $\Delta\phi$ between the signal and pump modes.
That is, the equations of motion are invariant under the transformation
$\hat{L}_0 \to \hat{L}_0 e^{i\Delta\phi}$, $\hat{L}_{2\vect{k}_p} \to
\hat{L}_{2\vect{k}_p} e^{-i\Delta\phi}$, using the notation introduced on
page~\pageref{eq:external-driving-hamiltonian}.
In an equilibrium condensate, the invariance of the energy under a
global rotation of phase implies the existence of a soft phase mode.
However, the existence of a free phase does not necessarily lead to
the same
consequences~\cite{wouters05,szymanska06:prl,wouters06,szymanska06:long},
as we will discuss next.

Amongst several distinctions between a laser and an equilibrium
condensate, one important difference is that for a laser, the
threshold condition is the balance of gain and
decay~\cite{haken70,haken75,scully97}, while for a condensate, the
Gross-Pitaevskii equation (gap equation, self-consistency condition)
involves the real parts of the self-energy.
This difference implies a difference in the spectrum of fluctuations.
Both cases have a pole in the fluctuation spectrum at $\omega=0$ and
zero wavevector, corresponding to global phase rotations.
However, for small, but non-zero wavevector, the spectrum first
acquires a real part for a true condensate, describing linearly
dispersing modes~\cite{khalatnikov,popov} (in an interacting system);
if instead one must balance gain and decay, the spectrum instead first
acquires an imaginary part, describing diffusive
modes~\cite{wouters05,szymanska06:prl,wouters06,szymanska06:long} (see
Fig.~\ref{fig:diffusive-poles}).
Although both the laser and strongly pumped polariton systems share
this diffusive structure, there is an important difference in the
properties of the phase mode between a conventional laser and the
strongly-pumped polariton system.
The difference is that, as discussed in
Sec.~\ref{sec:lasing-cond-superfl}, a conventional laser has a
discrete spectrum of wavevectors (all other modes rapidly decay),
while the microcavity polariton system has a continuum of in-plane
wavevectors with comparable decay rates.
The continuum of diffusive phase modes should thus lead to differences
between the coherence properties of a laser and a resonantly pumped
polariton system~\cite{wouters05,wouters06}.
The difference between diffusive and dispersive low energy degrees of
freedom may also have implications for pattern formation in nonlinear
systems~\cite{staliunas01,denz03}.

\subsubsection{Polarisation and spin relaxation}
\label{sec:polar-spin-relax}

By considering the spin of polaritons, with resonant pumping,
one can consider the coupled dynamics of the polarisations
of the pump, signal and idler modes.
This dynamics leads to a rich variety of physical effects, due to the
interplay between spin dependent stimulated scattering, and precession
induced both by momentum dependent TE-TM (Transverse-Electric,
Transverse-Magnetic) splitting, and other energy splittings due to
polariton-polariton interactions.
The following is a brief summary of the theories that have been
applied to explain these features.
Interest in the subject began with experiments, in both
c.w.~\cite{tartakovskii00} and pulsed~\cite{lagoudakis02} experiments,
that could not be explained by regarding each spin species
as acting independently.
In Ref.~\cite{tartakovskii00}, signal intensity as a function
of pump ellipticity (from linear to circular) was studied,
and a maximum found at an intermediate value.
In Ref.~\cite{lagoudakis02}, a large output signal was seen for a
linearly polarised pump, but a circularly polarised probe, and was 
explained there in terms of stimulated spin scattering.
In addition, the direction of the linear component of polarisation of
the signal was observed to vary as a function of the degree of
ellipticity of the pump beam.
This large rotation was discussed in Ref.~\cite{kavokin03} as a giant
Faraday effect, with the cavity amplifying the effect of spin
splitting of the exciton energy levels, with the spin splitting being due to
unequal spin populations of the exciton states.
In contrast, Ref.~\cite{eastham03:prb} showed that parts of the
above results could be explained by introducing coupling between
the two circular polarisations.
Such coupling provides two new terms; parametric scattering of cross
polarisations, and processes where pump and signal, or pump and idler
polaritons exchange polarisation.
These terms lead to a threshold power that depends on ellipticity of
the pump, and can under some conditions show lowest threshold for an
elliptic polarisation.

In a different experiment, Mart\'{i}n {\it et al.}~\cite{martin02}
investigated the dependence of the circular polarisation of the
non-linear emission on the detuning between the cavity photon modes
and excitons.
This result, showing oscillations of the polarisation, have been
attributed to the dependence of the TE-TM splitting on the
detuning~\cite{panzarini99}.
This TE-TM splitting, which depends on wave-vector, can lead to
time-dependent oscillations of the polarisation of the signal and
idler modes, as discussed in~\cite{kavokin04}.
In order to combine the precession due to such a splitting with a
spin-dependent stimulated scattering to the signal state, one is led
to write a spin dependent Boltzmann equation~\cite{kavokin04}.
This formalism was sufficient to reproduce the results of the
experiment described in Ref.~\cite{shelykh04:semiconductor}.
In that experiment the circular polarisation component of the signal
was studied as a function of pumping strength, for both linear and
circular polarisation.
It was found that, for a linear pump, there is a maximum of the degree
of circular signal polarisation near the nonlinear threshold.
As discussed in Ref.~\cite{shelykh04:semiconductor}, this effect can
be understood as a competition between two effects; self-induced
Larmor precession, which rotates the pseudospin describing the
polarisation from linear to circular polarisation, and stimulated
scattering to the ground state:
Far above threshold, the rate of scattering to the ground state is too
high to allow time for any change of polarisation in the pump mode.

An ingredient missing from the works listed so far was stimulated
scattering due to polariton-polariton interactions (as opposed to
polariton-phonon scattering, stimulated by final state polariton
population).
Shelykh {\it et al.}\ considered the dominant, parallel spin,
interaction in Ref.~\cite{shelykh04:spin}.
However it was later realised~\cite{kavokin05,glazov05} that the
scattering of anti-parallel spin states, though small, is important,
as it leads (at high densities) to a $90^{\circ}$ rotation of linear
polarisation direction between pump and signal.
In a recent experiment~\cite{krizhanovskii06}, the polarisation of the
output signal was studied for a linearly polarised pump, as a function
of pump power and angle of the linear polarisation.
This experiment showed that, just above threshold, the signal was
elliptical, with the major axis (the direction of the linear
component) rotated with respect to the linear polarisation of the pump
--- i.e.\ the direction of major axis depended on the direction of
polarisation of the pump, but the angle between the two directions was
not constant.
In addition, the degree of linear polarisation of the signal decreased
far above threshold (as well as the degree of circular polarisation as
observed in Ref.~\cite{shelykh04:semiconductor}).
This rotation was reproduced in a model combining precession due to
static and self-induced (Larmor) splittings (including an extra
in-plane splitting, as discussed in Ref.~\cite{krizhanovskii06}) as
well as spin rotation in stimulated polariton-polariton scattering.
The reduction of linear polarisation far above threshold was explained
by rapid self-induced Larmor precession, which rotates the linear
polarisation direction so rapidly that it averages to zero.

\section{Connection to other systems, conclusions}
\label{sec:conn-other-syst}

\subsection{Atomic gases and Feshbach resonances}
\label{sec:feshb-reson-ferm}

Microcavity polaritons and their condensation are related to the
physics of two-component atomic Fermi gases near Feshbach resonances.
In particular, the crossover from a fluctuation dominated phase
boundary to a mean-field phase boundary with increasing polariton
density is closely related to the BEC-BCS crossover recently studied
in these atomic gases~\cite{Regal04,Zwierlein04}.
For atomic gases, in contrast to polariton experiments, the density of
particles is typically kept fixed, while the interaction strength is
varied via magnetically tunable Feshbach resonances, allowing one to
go from a BEC of bound molecules to a BCS state of fermionic pairs.
The interaction strength is tuned by changing the detuning between the
zero of energy for pairs of atoms in their original spin states, and a
closed channel resonance level of atoms in some higher energy spin
configuration  (See Ref~\cite{koehler06} and Refs therein for more details).

The Bose-Fermi model used for polaritons and described in
section~\ref{sec:boson-ferm-gener} is very similar to the model
initially proposed for the description of the BEC-BCS crossover in
atomic Fermi gases~\cite{Timmermans01}.
The fermionic operators $b^{\dagger},a^{\dagger}$ in the polariton
model, introduced in Eq.~(\ref{eq:dicke-model-fermions}), are
analogous to the two spin species of atoms in the Feshbach resonance
model; the photon is analogous to the closed channel resonance level;
the dipole interaction relates to the hyperfine interaction; and
the polariton to the Feshbach molecule.
There are, however, a number of important differences between the two
models; most notably the absence of a direct four-fermion interaction
in the polariton model,  and the existence of an energy
dependence, and a distribution, of exciton-photon coupling strengths.
Secondly there is a marked difference of mass ratios:
For polaritons, the photon mass is typically a factor of $10^5$ times
smaller than the exciton mass; in the Feshbach case the closed channel
resonance has a dispersion controlled by a mass which is twice the
atomic mass.
Note also that there is a difference in interpretation between the
photon and the  closed channel resonance level in Feshbach
resonance.
Although inside the microcavity a photon is not an eigenstate --- the
upper and lower polaritons are instead eigenstates --- outside the
cavity it is possible to physically separate the photons from
excitons.
Similarly, the closed channel resonance level is not an eigenstate;
however, in distinction from the photon, it cannot be physically
separated from the other two-body states to which it is coupled:
In general one cannot have a given molecular resonance level in
isolation from other molecular states.
Another difference between the systems comes in how the gap equation
[i.e.\ Eq.~(\ref{eq:gpe-fermi})], and constraint on the density can be
used to find chemical potential and temperature:
In the mean-field (BCS) limit of the atomic case, the closed channel
resonance level lies at high energies, the chemical potential lies
well within the band of fermionic states, and so the density depends
only on chemical potential, and not on temperature.
In the polariton model, the chemical potential remains below the
bottom of the band of fermionic states, and so both temperature and
chemical potential influence the density.
Rather, the fluctuation dominated (BEC) to mean-field (BCS) crossover
of the polariton model
is in the nature of which excitations are responsible for depopulating
the condensate at finite temperatures.

Another issue worth considering is that further analysis has
questioned the need to use the Bose-Fermi model in applications to
experimentally relevant Feshbach resonances.
It has been shown~\cite{Koehler03:PRL,Koehler03:PRA} that the molecules
created in the vicinity of a Feshbach resonance are halo dimers
extending over large distances in which the closed channel admixture
is tiny.
Thus, the resonance level acts to enhance the effective interaction
between fermionic pairs, but the crossover to loosely bound molecules
does not rely on the macroscopic occupation of this level (as was
initially suggested).
Since the resonance level is not significantly occupied, it was
shown, by using realistic atomic potentials~\cite{Szymanska05:PRA},
that the BCS-BEC crossover in atomic Fermi gases near Feshbach
resonances is of the same nature as originally considered by
Leggett~\cite{Leggett80}.
That is, it is based on the smooth crossover of the pair size and not
on the macroscopic occupation of the resonance level.
Thus, microcavity polariton experiments in the normal operating
conditions of large photon field occupation present the first
experimental realisation of the BEC-BCS crossover which differs
substantially from the original
scenario~\cite{Leggett80,keeling04:polariton,keeling05}.

Finally, it is interesting to note that the same idea of parametric
scattering and amplification in a resonant pumping configuration as
is described in Sec.~\ref{sec:resonant-pumping}, has been recently
applied to an ultracold dilute gas of bosonic atoms confined in an
optical lattice~\cite{campbell06} (see also the related
  experiments on dynamic instabilities~\cite{burger01,fallani04}, and
  four wave mixing of matter waves~\cite{deng99}).
There, a ${}^{87}$Rb Bose-Einstein condensate was loaded into a moving
one-dimensional optical lattice.
The optical lattice causes the atomic dispersion to deviate from
quadratic, and allows parametric scattering: Atom pairs with initial
momentum $k_p$ inherited from the moving lattice scatter elastically
into two final states; $k_p-k$ and $k_p+k$.
By generating a seed of atoms with momentum $k_p-k$, parametric
amplification of both the seed and the conjugate momentum has been
observed, with a gain determined by the atomic scattering length.

\subsection{Excitons, quantum hall bilayers, triplons}
\label{sec:excit-cond-double}

The high quantum degeneracy temperatures and the high degree of
control obtained by laser photo-excitation suggest that excitonic
systems should also provide excellent environments in which to study
macroscopic coherence phenomena.
Much interest over the last two decades has been attracted by excitons
in bulk Cu${}_2{}$O crystals (see, e.g.,Ref.~\cite{jang06} and references
therein).
Excitons in Cu${}_2{}$O have a large binding energy ($\sim 150$meV)
and the fact that the direct dipole transition from the exciton ground
state is forbidden guarantees low radiative recombination rates.
In Cu${}_2{}$O crystals however non-radiative recombination processes
such as the Auger effect cause loss and heating, and represent the main
obstacle to the observation of quantum degeneracy in these structures.

Low radiative recombination rates together with high cooling rates can
be obtained for spatially indirect excitons in coupled quantum wells.
Here, an electric field is applied along the growth direction, in such
a way that electrons and holes separate into different wells.
In contrast to the case of direct excitons, lifetimes up to few
microseconds can be achieved, while the high cooling rate gives a much
shorter thermalisation time, typically in the nanosecond range.
As a consequence, in coupled quantum well structures, thermal
equilibrium with the lattice can be relatively easily obtained by
either waiting a few nanoseconds after photo-excitation or allowing
the excitons to travel away from the excitation spot.
While much experimental work has been done on these structures in the
   last few years\cite{butov02natureA,butov02natureB,snoke02nature,
   snoke05:PRL} (for review see \cite{butov04,snoke02:science}),
   unambiguous evidence for Bose-Einstein condensation of indirect
   excitons in coupled quantum wells is still missing and is the
   subject of intensive on-going studies.
Recently, gases of indirect excitons have been trapped and
equilibrated using in-plane potentials, either by applying localised
stress to change the local band energies~\cite{voros06:PRL} or by
means of optical~\cite{hammack06:PRL} or
electrostatic~\cite{chen06:PRB} traps.
The confinement of indirect excitons prevents the fast reduction of
initial density which occurs in the absence of trapping due to fast
expansion driven by their strong dipole-dipole repulsive interaction
and their relatively high mobility.

It is interesting to note that, for an untrapped system, while the
formation of an external excitonic ring has been explained as an
in-plane charge separation, at low temperatures it has been shown
that, such a ring can separate into a periodic array of beads, and that
the light emitted by each of these beads is
coherent~\cite{yang06:PRL}.
The origin of such a phenomenon is still unknown.

Recently, experimental evidence suggests that, under appropriate
conditions, an electron-electron semiconductor bilayer system in the
Quantum Hall regime can condense into a superfluid state which might
be interpreted as an excitonic-like
superfluid~\cite{eisenstein04:nature}.
In a bilayer 2D electron system with total filling factor
$\nu_{T}=1$, excitonic pairs can be thought of as formed by filled
electron states in one layer and empty electron states in the second
layer.
By changing the ratio between intra- and inter-layer Coulomb
interactions, signatures of the transition to a condensed excitonic
phase have been shown by a dramatic increase in the tunnelling rate
between the two layers at zero interlayer
voltage~\cite{spielman00:PRL}, in Coulomb drag
measurements~\cite{kellog02:PRL}, and in counterflow
measurements~\cite{tutuc04:PRL}.

Another condensed matter system in which a phase transition can be
described as condensation of an excitonic mode is that of magnetic
``triplon'' excitations. 
This has been seen in a variety of compounds including
TlCuCl$_3$~\cite{nikuni00,ruegg03}, BaCuSi$_2$O$_6$~\cite{jaime04} and
Pb$_2$V$_3$0~\cite{waki04}, where by changing the applied magnetic
field, there is a crossing between spin singlet and spin triplet
excitations.
The resultant magnetic phase transition can be described as condensation
of the triplet magnon mode.
Very recently such a transition was seen at room temperature, using
parametric laser pumping to create a non-equilibrium density of
triplons in the compound Y$_3$Fe$_2$(FeO$_4$)$_3$~\cite{demokritov06}.

\subsection{Conclusions}
\label{sec:conclusions}

In this article we reviewed the experimental results to date
demonstrating coherence in microcavities, and discussed the variety of
theoretical models and techniques that have been used to describe it.
We discussed experiments with both non-resonant pumping, in which
coherence may spontaneously arise from an initially incoherent source
of polaritons, and the optical parametric amplifier and optical
parametric oscillator:  
Both kinds of experiments allow one to explore the interplay of
strong-coupling, coherence, lasing and condensation.
In our discussion of theoretical descriptions, we highlighted those
aspects of these solid state systems which can introduce new questions
about coherence --- disorder, decoherence, particle flux, potentially
non-thermal distributions.

Of course, there are subjects connected to coherence in microcavities
that we have not had space to discuss, or have only discussed briefly.
For example, we have only briefly mentioned here questions about the
dynamics of condensate formation, and of the polariton response
following short pump pulses.
The last years have seen rapid experimental progress in this field,
with the first convincing evidence of coherence developing from
incoherently injected polaritons in a variety of
systems~\cite{kasprzak06:nature,snoke06:condensation,baumberg06}.
This experimental progress both gives hope for the possibility of
future experiments, and applications on coherent microcavity
polaritons, as well as focusing attention on those areas in which
further theoretical work is necessary.
There are areas in which our discussion has been brief because some
questions have only been partially addressed to date:
Questions about hydrodynamics in such partially coherent, pumped
decaying systems.
In order to address questions about the generic behaviour of such
systems, it is important to understand how the variety of models used
to describe polaritons relate, and what features each can explain.
Finally, let us mention that there remains an interesting topic which
can be the subject of much future research: what new experiments are
possible in these light-matter systems that were not possible in
either lasers or in atomic gases.

\ack We are grateful to L.~S.~Dang, P.~R.~Eastham, J.~Kasprzak,
B.~D.~Simons for helpful discussions, and to I.~Carusotto, A.~Kavokin,
T.~K{\"o}hler, D.~Sarchi, V.~Savona, L. Vi\~{n}a and R. Zimmermann for
critical reading of, and helpful suggestions on the manuscript.
F.M.M. and M.H.S.  would like to acknowledge financial support from
EPSRC. J.K. would like to acknowledge financial support from the
Lindemann Trust and Pembroke College Cambridge. This work is supported
by the EU Network ``Photon mediated phenomena in semiconductor
nanostructures'' HPRN-CT-2002-00298.

\section*{References}
\newcommand\textdot{\.}

\end{document}